\documentclass[abstract]{B20924+30}
\pdfoutput=1
\usepackage[lofdepth,lotdepth]{subfig}
\usepackage{graphicx}
\usepackage{txfonts}
\usepackage{natbib}
\usepackage{float}

\newcommand{\mJybeam}{mJy beam$^{-1}$}

\newcommand{\kms}{$\,$km$\,$s$^{-1}$}
\newcommand{\whz}{$\,$W$\,$Hz$^{-1}$}
\newcommand\addtag{\refstepcounter{equation}\tag{\theequation}}

\bibpunct{(}{)}{;}{a}{}{,}

\begin{document}

\title{Radiative age mapping of the remnant radio galaxy B2~0924+30: the LOFAR perspective\thanks{The LOFAR and WSRT images used to derive the spectral index and ageing maps are available in electronic form at the CDS via anonymous ftp to cdsarc.u-strasbg.fr (130.79.128.5) or via http://cdsweb.u-strasbg.fr/cgi-bin/qcat?J/A+A/}}

\titlerunning{LOFAR studies of B2~0924+30}
\authorrunning{Shulevski et al.}	
\author{A.~Shulevski\inst{1,2}\and
	R.~Morganti\inst{1,2}\and
	J.~J.~Harwood\inst{1}\and 
	P.~D.~Barthel\inst{2}\and
	M.~Jamrozy\inst{3}\and
	M.~Brienza\inst{1,2}\and
	G.~Brunetti\inst{4}\and
	H.~J.~A.~R\"{o}ttgering\inst{5}\and
	M.~Murgia\inst{6}\and
	G.~J.~White\inst{7,8}\and
	J.~H.~Croston\inst{9}\and
	M.~Br\"{u}ggen\inst{10}
	}
	\institute{ASTRON, the Netherlands Institute for Radio Astronomy, Postbus 2, 7990 AA, Dwingeloo, The Netherlands\\
		   \email{shulevski@astron.nl}
		   \and
		   University of Groningen, Kapteyn Astronomical Institute, Landleven 12, 9747 AD Groningen, The Netherlands\and
		   Obserwatorium Astronomiczne, Uniwersytet Jagiello\'{n}ski, ul Orla 171, 30-244, Krak\'{o}w, Poland
		   \and
		   IRA-INAF, via P. Gobetti 101, 40129 Bologna, Italy
		   \and
		   Leiden Observatory, Leiden University, Niels Bohrweg 2, 2333 CA Leiden, The Netherlands
		   \and
		   INAF-Osservatorio Astronomico di Cagliari, Via della Scienza 5, I-09047 Selargius (CA), Italy
		   \and
		   Department of Physics and Astronomy, The Open University, Walton Hall, Milton Keynes MK7 6AA, England\and
		   RAL Space, The Rutherford Appleton Laboratory, Space Science and Technology Department, Chilton, Didcot, Oxfordshire OX11 0QX, England\and
		   School of Physics and Astronomy, University of Southampton, Southampton SO17 1BJ, UK\and
		   University of Hamburg, Gojenbergsweg 112, 21029 Hamburg, Germany
		  }
\date{Received 3 November 2016 / Accepted 6 January 2017}

\abstract{We have observed the steep spectrum radio source B2~0924+30 using the LOw Frequency ARray (LOFAR) telescope. Hosted by a z=0.026 elliptical galaxy, it has a relatively large angular size of 12\arcmin\ (corresponding to 360 kpc projected linear size) and a morphology reminiscent of a remnant Fanaroff-Riley type II (FRII) radio galaxy. Studying active galactic nuclei (AGN) radio remnants can give us insight into the time-scales involved into the episodic gas accretion by AGNs and their dependence on the AGN host environment.
The proximity of the radio galaxy allows us to make detailed studies of its radio structure and map its spectral index and radiative age distribution. We combine LOFAR and archival images to study the spectral properties at a spatial resolution of 1\arcmin\ .
We derive low frequency spectral index maps and use synchrotron ageing models to infer ages for different regions of the source. Thus, we are able to extend the spectral ageing studies into a hitherto unexplored frequency band, adding more robustness to our results.
Our detailed spectral index mapping, while agreeing with earlier lower resolution studies, shows flattening of the spectral index towards the outer edges of the lobes. The spectral index of the lobes is $ \alpha_{140}^{609} \sim -1 $ and gradually steepens to $ \alpha_{140}^{609} \sim -1.8 $ moving towards the inner edges of the lobes. Using radiative ageing model fitting we show that the AGN activity ceased around 50 Myr ago. We note that the outer regions of the lobes are younger than the inner regions which is interpreted as a sign that those regions are remnant hotspots.
We demonstrate the usefulness of maps of AGN radio remnants taken at low frequencies and suggest caution over the interpretation of spectral ages derived from integrated flux density measurements versus age mapping.
The spectral index properties as well as the derived ages of B2~0924+30 are consistent with it being an FRII AGN radio remnant. LOFAR data are proving to be instrumental in extending our studies to the lowest radio frequencies and enabling analysis of the oldest source regions.}


\maketitle

\section{Introduction}
\label{c5:intro}

Although active galactic nuclei (AGN) have been observed to influence their surrounding interstellar and intergalactic medium \citep[ISM/IGM,][]{McNamara2012, Randall2010}, the impact this may have depends on a number of relatively poorly known factors, in particular the duty-cycle of the activity, i.e. the portion of time the super massive black hole (SMBH) is active \citep{Mendygral2012}.

Tracers of past AGN accretion episodes can be observed at radio wavelengths. In the case of radio-loud AGN, their ages (and duty cycle in the case of restarted sources) can be derived using the spectral properties of the radio plasma. Once the accretion of matter onto its SMBH stops, the ejection of plasma jets ceases, terminating the supply of fresh electrons into the radio lobes. These synchrotron radio remnant regions then slowly fade as time passes owing to preferential cooling of high energy particles and/or adiabatic expansion. Consequently, their spectral index steepens ($ \alpha < -1 $)\footnote{We define the spectral index as: $ S \, \sim \, \nu^{\alpha} $}, and breaks appear in the radio spectrum \citep{RefWorks:126, RefWorks:187, RefWorks:125, RefWorks:188}. If the radio emission restarts (as observed in a number of cases) this would further modify the shape of the source's radio spectrum \citep{RefWorks:34} and may influence the source's morphology.

Thus, radio studies enable us to identify the presence and the timescales of this type of cycle of activity. Selecting radio sources that have a steep spectral index over a range of frequencies is the predominant way of discovering AGN radio remnants\footnote{To distinguish radio sources produced by past AGN activity, as opposed to steep spectrum sources found in galaxy clusters (relic, phoenix) we name the former AGN remnants.}, i.e. sources in which the radio source has switched off. However, the question remains as to why there are so few remnants detected (a few dozen in total) relative to the entire population of active radio galaxies.

Because of the steepness of the spectrum, the low radio frequency observational window is the one where remnants can be more easily detected. For this reason, the recent availability of new deep images from the low frequency array \citep{RefWorks:157} has revamped the search and the study of these objects. The first searches with LOFAR have already provided promising results with the percentage of remnants ranging between 10\% and 30\% \citep{Brienza2016b, Hardcastle2016}. These observations are starting to put constraints on radio galaxy evolution models and are helping us to understand what the relevant physical processes involved in remnant evolution are.

A prerequisite for the complete characterization of the AGN duty cycle is the determination of the active and switched off times of individual objects over a statistically significant sample.
In studies of several double-double radio galaxies (DDRGs), \cite{RefWorks:178} and \cite{Orru2015} find that they have a relatively rapid duty-cycle with the time elapsed between the periods of activity being a fraction of the total age of the source.
Bona fide remnants can show a total age of over a hundred Myr \citep{Harris1993, Venturi1998, RefWorks:206, RefWorks:250}. The duration of their remnant phase in most cases appears to be shorter or comparable to that of the active phase \citep[][respectively]{RefWorks:96, RefWorks:34, RefWorks:2}. Cases where the remnant phase is (much) longer than the active phase are, so far, rarer \citep[e.g.][]{Brienza2016a}.
The duty cycle likely has a dependence on galaxy mass and source power during the active phase, as suggested by statistical studies \citep{RefWorks:41, RefWorks:43}.

Detailed remnant studies have so far been limited to just a few cases and have often not been carried out at sufficiently high spatial resolution to enable the investigation of the radiative ages across the sources, and of their activity histories.

\subsection{B2~0924+30}
\label{c5:intro:target}

An object that offers this possibility is B2~0924+30, the target of this paper. Its host, IC~2476 (UGC~5043), is the brightest member of the relatively poor Zwicky cluster 0926.5+30.26 \citep{RefWorks:97, RefWorks:16, White1999}, which is located at a redshift of $ z = 0.026141 $. Its Sloan Digital Sky Survey \citep[SDSS;][]{RefWorks:240} spectrum does not show emission lines indicative of an optical AGN.
The radio luminosity\footnote{The adopted cosmology in this work is: $  H_{0} = {73} $ \kms Mpc$^{-1}$ , $ \Omega_{matter} =  0.27 $, $ \Omega_{vacuum}  =  0.73$. At the redshift of B2~0924+30, $1^{\arcsec} =  0.505 $ kpc; its luminosity distance is 109.6 Mpc \citep{RefWorks:155}.} of B2~0924+30 is $ L_{\mathrm{1400MHz}} \sim 10^{23.8} $ \whz . It lacks a discernible radio core or jets / hotspots and is considered to be an AGN remnant by \cite{RefWorks:97}. Spectral index studies by \cite{RefWorks:206} show that the spectral index steepens going from the lobes to the inner regions, and the overall spectral index distribution is steeper ($ \alpha \, \sim \, -1 $) than that observed in most active radio galaxies.

\cite{RefWorks:206} have also performed a radiative ageing analysis of B2~0924+30 and find an overall average source age of $ 54_{-11}^{+12} $ Myr.

We expand on previous research efforts by extending the spectral index studies to even lower radio frequencies. Using LOFAR  we have derived the highest spatial resolution spectral index mapping to date extending to 140 MHz, enabling us to characterize in detail the spectral properties of the remnant lobes. Our aim is also to perform a resolved radiative age mapping of the source to better ascertain its activity history.

The organization of this paper is as follows. Section \ref{c5:obsdata} describes the data used in this study and outlines the data reduction procedure. Section \ref{c5:res} outlines our results; in Sect. \ref{c5:spec_an} we present the spectral analysis results and we discuss the derived source ages in Sect. \ref{c5:rad_ages}. We discuss the implications of our study in Sect. \ref{c5:disc}.

\section{Observations and data reduction}
\label{c5:obsdata}

The target was observed with the LOFAR high band antennas (HBA) on the night of March 13, 2014, for a total on source time of 7.5 hours. The observations were obtained in the interleaved mode, using the full Dutch array of 38 antenna stations. The two HBA antenna fields of each of the core stations were treated as separate stations and of the HBA fields of the remote stations only the inner tiles were used (this configuration is known as HBA\_DUAL\_INNER). 3C~196 was observed as a flux calibrator source for two minutes, followed by a scan of the target of 30 minutes duration with a one minute gap between calibrator and target scans that allowed for beam forming and target re-acquisition. We recorded 325 sub-bands (SBs), over the 63.5 MHz of bandwidth between 116 MHz and 180 MHz. Each SB has 64 frequency channels and a bandwidth of 195.3 kHz. The integration time was set to 2 seconds for both calibrator and target. Four polarizations were recorded. The HBA station field of view (FoV, primary beam)  covers around 5 degrees full width at half maximum (FWHM) at 140 MHz. The station beams are complex valued, time, frequency and direction dependent, and are not the same for all of the stations.

\begin{table}[!htpb]
\noindent \caption{\small LOFAR HBA data properties}
\label{table:1}
\small
\begin{tabular}{ p{4cm} p{4cm}}
\hline\hline\\
Channels per SB (192 kHz) & 64 \\
Central Frequency & 150 MHz \\
Bandwidth & 63.5 MHz \\
Integration time & 2 s\\
Observation duration & 7.5 h\\
Polarization & Full Stokes \\
UV coverage & $ 0.1 $k$\lambda - 20 $k$\lambda $\\
\hline
\end{tabular}
\end{table}

The data were pre-processed by the observatory pipeline \citep{RefWorks:180} as described below. Each SB was automatically flagged for radio frequency interference (RFI) using the AOFlagger \citep{RefWorks:133}, and then averaged in time to 10 seconds per sample and in frequency by a factor of 16, making the frequency resolution of the output data 4 channels / SB. The calibrator data were used to derive amplitude solutions for each (Dutch) station using the Blackboard self calibration \citep[BBS,][]{RefWorks:182} tool that takes into account the LOFAR station beams variation with time and frequency. The flux density scale of \cite{RefWorks:181} was used in the calibration model for 3C~196 ($ S_{150} \, = \, 83 $ Jy).

The amplitudes of the target visibilities were corrected using the derived calibrator solutions. The target visibilities were then phase-(self)calibrated incrementally, using progressively longer baselines to obtain the final (highest) image resolution. The initial phase calibration model was derived from the VLSS\footnote{VLSS is the VLA Low frequency Sky Survey carried out at 74 MHz \citep{RefWorks:128}} catalogue that covers the FoV out to the first null of the station beam, which contains spectral index information for each source in the model. Before initializing the calibration, we concatenated the data into 4 MHz (20 SB) groups previously averaging each SB to 1 frequency channel to reduce the data size. We chose this set-up to maximize the S/N while maintaining frequency-dependent ionospheric phase rotation to a manageable level. In the calibration, we neglected direction-dependent effects (ionosphere and residual clock errors on longer baselines). However, since our target is in the phase center of the FoV, these issues do not represent a limit to our science goals (as demonstrated below).

The imaging was performed using the LOFAR AW imager \citep{RefWorks:183}, which incorporates the LOFAR beam and uses the A-projection \citep{RefWorks:186} algorithm to image the entire FoV. We used Briggs \citep{RefWorks:185} weights with the robustness parameter set to $ 0 $, and imaged by selecting baselines larger than 0.1 $ k\lambda $. Ten self-calibration steps were performed, each using a sky model generated in the previous cycle and each subsequent one using larger baseline lengths.
The self-calibration resulted in images that cover the HBA band, out of which we selected a low- and a high-resolution one (only images not affected by calibration errors)\footnote{The selection resulted in six high-resolution and six low-resolution images}.

\begin{table}[!htpb]
\noindent \caption{\small Image properties}
\label{c5:table:2}
\small
\begin{tabular}{c c c c}
\hline\hline\\
\small ID & \small $ \nu $ [MHz] & \small $ \sigma $ [mJy/b] & Beam size \\
\hline\\
LOFAR\tablefootmark{a} & 113 & 8.3 $ \vert $ 4.5 & $ 56\farcs6 \times 40\farcs9 \, \vert \, 20\farcs2 \times 14\farcs1 $ \\
LOFAR\tablefootmark{a} & 132 & 4.3 $ \vert $ 3.1 & $ 48\arcsec \times 35\farcs4 \, \vert \, 22\arcsec \times 16\farcs7 $ \\
LOFAR\tablefootmark{a} & 136 & 4.3 $ \vert $ 3 & $ 46\farcs9 \times 34\farcs3 \, \vert \, 21\farcs7 \times 17\farcs1 $ \\
LOFAR\tablefootmark{a} & 160 & 2.3 $ \vert $ 1.9 & $ 51\farcs9 \times 37\farcs6 \, \vert \, 20\arcsec \times 17\farcs9 $ \\
LOFAR\tablefootmark{a} & 163 & 2 $ \vert $ 1.8 & $ 56\farcs6 \times 38\farcs2 \, \vert \, 20\farcs2 \times 17\farcs8 $ \\
LOFAR\tablefootmark{a} & 167 & 1.8 $ \vert $ 1.7 & $ 51\farcs1 \times 37\farcs5 \, \vert \, 20\farcs5 \times 17\farcs5 $ \\\\
LOFAR\tablefootmark{a}\tablefootmark{c} & 140 & 2.5 $ \vert $ 1.2 & $ 60\arcsec \times 43\farcs5 \, \vert \, 22\arcsec $ \\\\
WENSS & 325 & 3.6 & $ 54\arcsec \times 108\arcsec $ \\
WSRT\tablefootmark{b} & 609 & 0.77 & $ 29\arcsec \times 56\arcsec $ \\
NVSS & 1400 & 0.45 & 45\arcsec \\
\hline
\end{tabular}
\tablefoot{
The image noise and beam size are given for the low and high resolution images respectively.\\
\tablefoottext{a}{This work}
\tablefoottext{b}{\cite{RefWorks:206}}
\tablefoottext{c}{Averaged image}
}
\end{table}

We smoothed the high- and low-resolution LOFAR image sets to an identical restoring beam size and averaged them to obtain two averaged images, each having a bandwidth of 28 MHz. We used these images for morphological studies of the target source. The smoothed, individual images were used in our ageing analysis. Table \ref{c5:table:2} lists the image properties for the LOFAR image set, as well as survey (WENSS\footnote{WENSS is the Westerbork Northern Sky Survey carried out at 325 MHz \citep{RefWorks:138}} and NVSS\footnote{NVSS stands for the NRAO VLA Sky Survey carried out at a frequency of 1400 MHz \citep{RefWorks:139}}) images used in our subsequent analysis.

\begin{figure}[!htpb]
\centering
\subfloat[][HBA - Low resolution]{\includegraphics[width=0.5\textwidth]{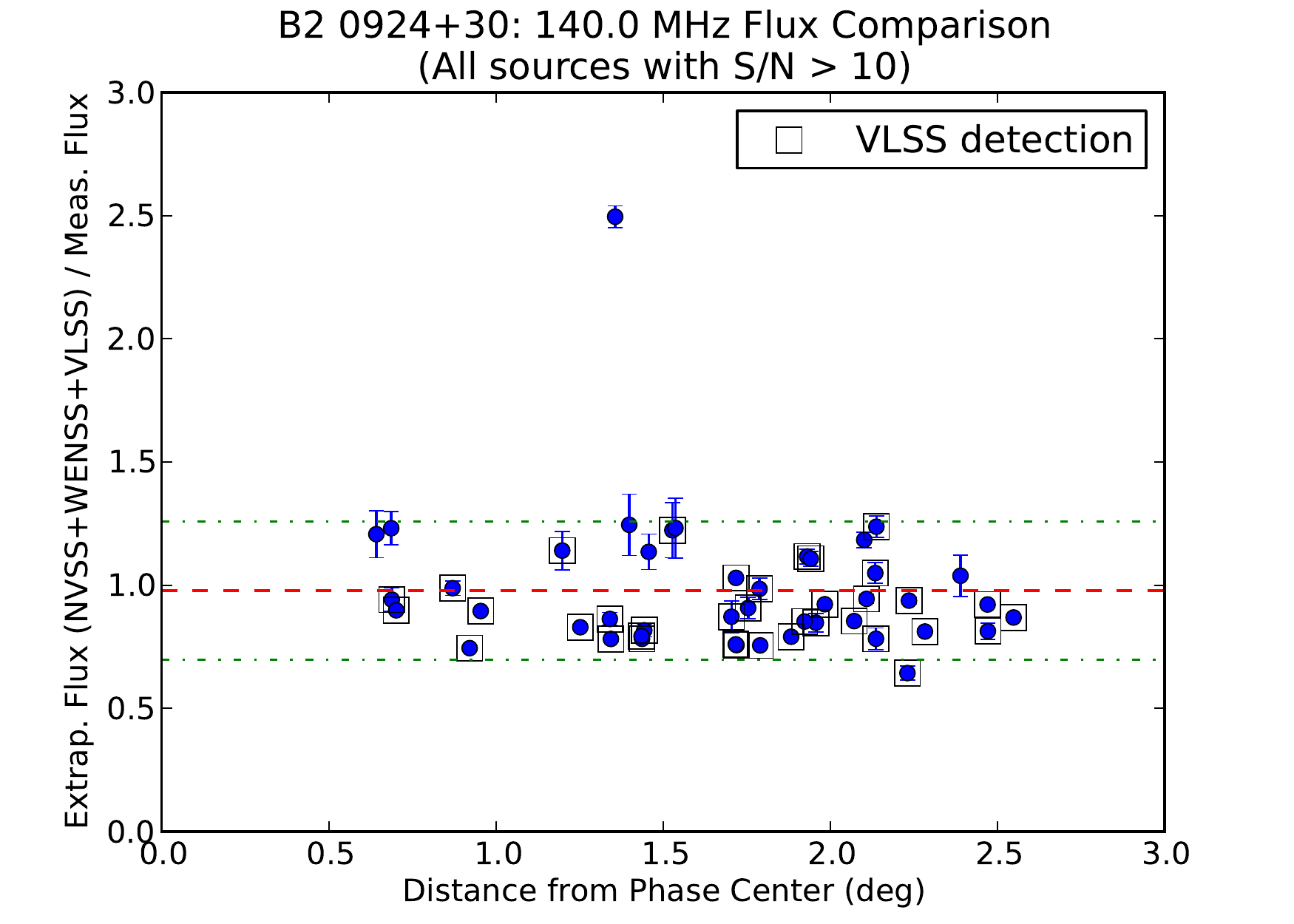}
\label{c5:FoVfluxes:subfig2}}
\quad
\subfloat[][HBA - High resolution]{\includegraphics[width=0.5\textwidth]{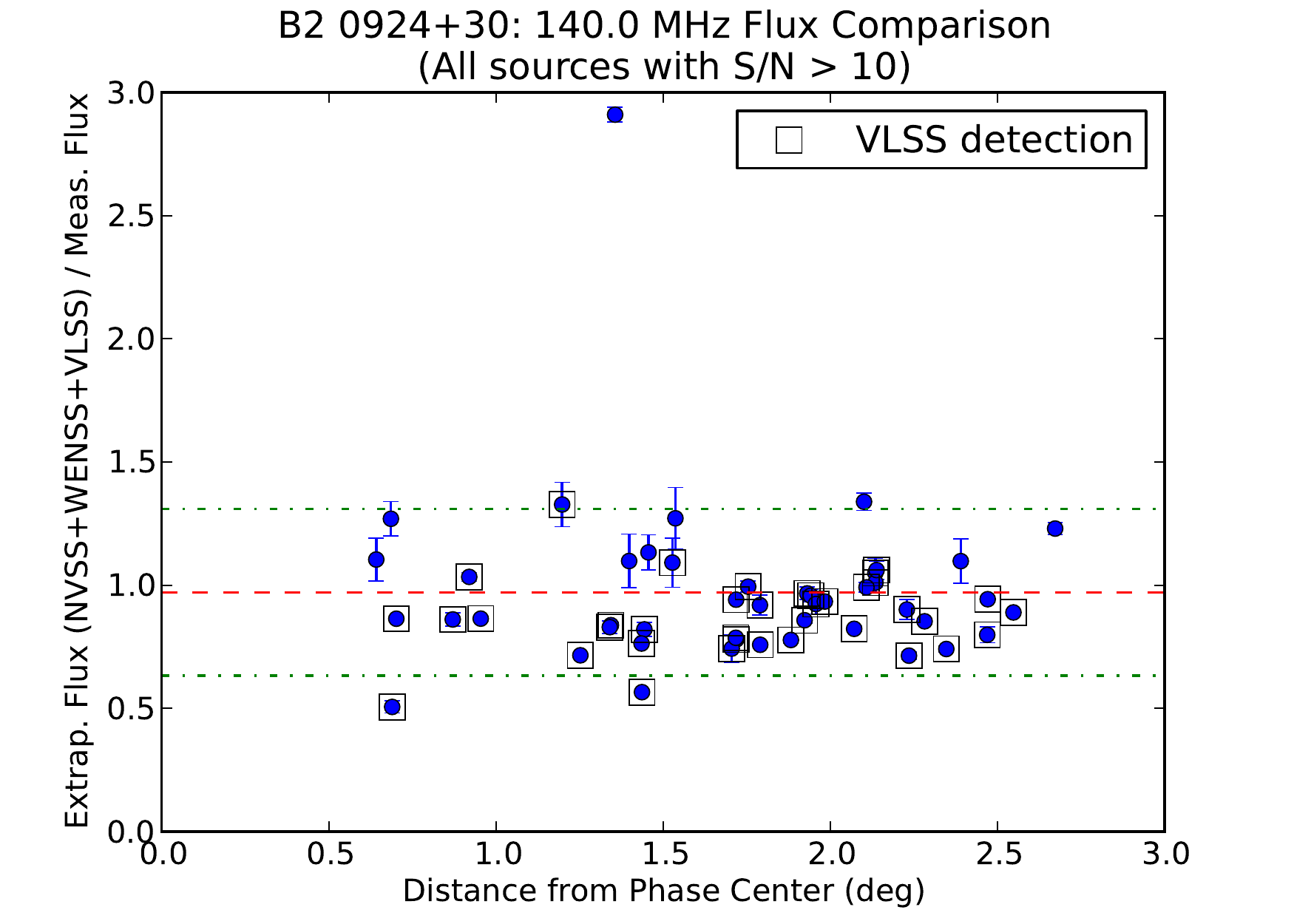}
\label{c5:FoVfluxes:subfig3}}
\caption{Ratio of measured and catalogue extrapolated flux densities for our high- and low-resolution averaged LOFAR images.}
\label{c5:FoVfluxes}
\end{figure}

To check whether the station beam correction applied by the AW imager resulted in correct flux-density scaling across the FoV, we have the PyBDSM source finder package \citep{pybdsm} and have extracted point sources from our averaged images. Then, we matched the extracted sources with survey catalogs (VLSS, WENSS and NVSS using a $ 30 $\arcsec\ match radius) and determined the catalogue flux density for each source by interpolating the flux densities from the catalogue entries to the LOFAR frequency. Finally, we divided the obtained catalogue flux density at 140 MHz with the measured flux density from the LOFAR image. Assuming power-law spectra, the ratio should be unity if the station beam correction gives correct fluxes over the FoV. The results are given in Fig. \ref{c5:FoVfluxes}. We can see that for both of the HBA images the points cluster around $ 1 $, which shows that the flux correction over the FoV applied by the AW imager gives reasonable flux-density values. The scatter is around $ 20 $\%.

Owing to an incomplete HBA beam model, the influence of the grating side-lobes was not properly taken into account during processing, which results in a systematic bias in the measured source fluxes across the LOFAR band. This shows up as a systematic steepening of the in-band spectral index. We applied a LOFAR beam normalization correction factor to the measured flux densities to mitigate the effect.

\section{Results}
\label{c5:res}

\begin{figure*}[!ht]
\centering
\includegraphics[width=1.0\textwidth]{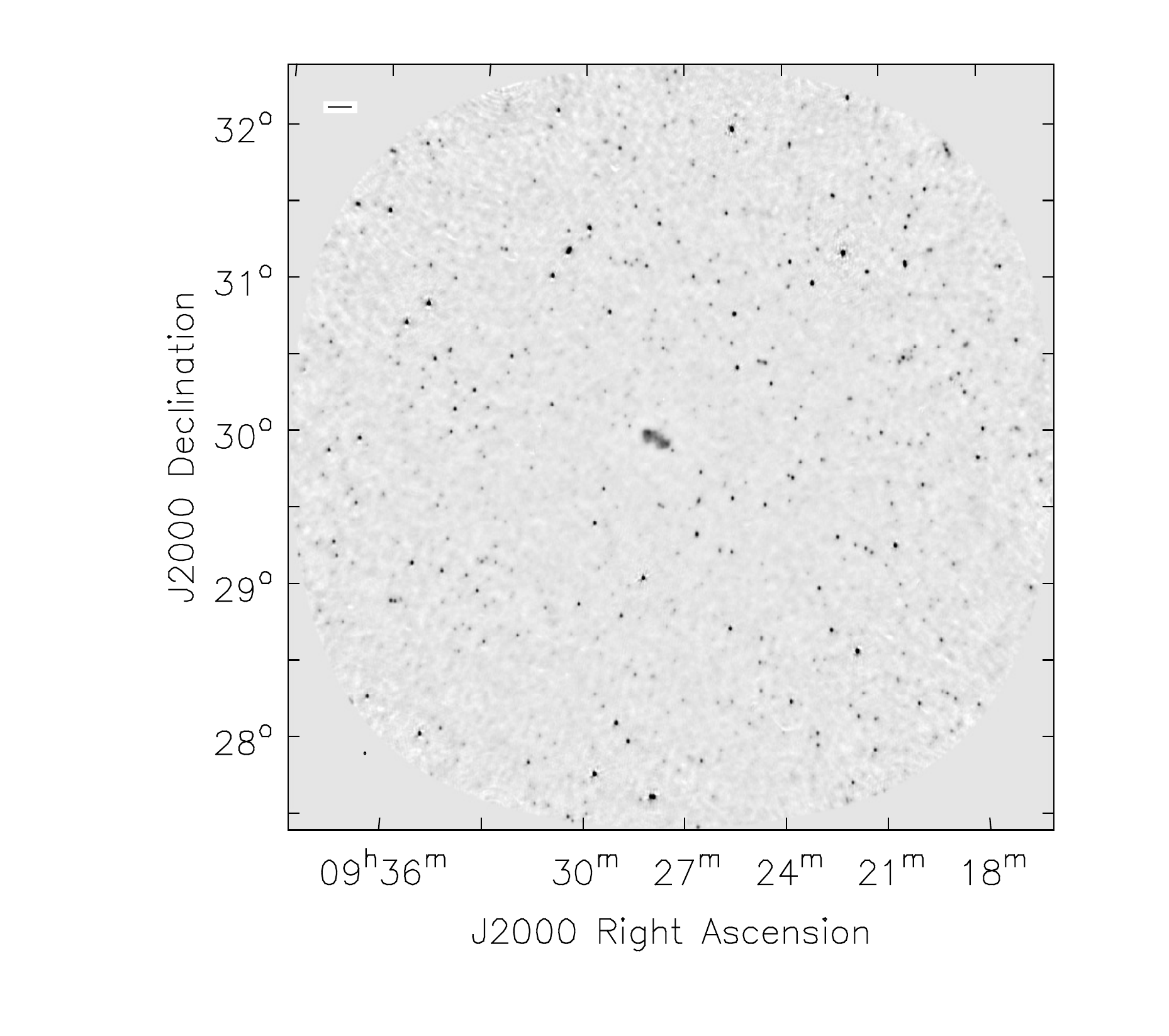}
\caption{LOFAR FoV, centred on B2~0924+30, low-resolution image averaged over a bandwidth of 28 MHz. Beam size: $ 60\arcsec \times 43\farcs5 $, $ \sigma \, = \, 2.5 $ \mJybeam\ .}
\label{c5:fullFoV}
\end{figure*}

Figure \ref{c5:fullFoV} shows the low-resolution $ 5^{\circ} \times 5^{\circ} $ LOFAR image obtained by smoothing and averaging together six images taken across the LOFAR band (listed in Table \ref{c5:table:2}). It has a resolution of $ 60\arcsec \times 43\farcs5 $ and an r.m.s. noise level of $ 2.5 $\mJybeam.

\begin{figure*}[ht!]
\centering
\subfloat[][B2~0924+30 LOFAR HBA]{\includegraphics[width=0.5\textwidth]{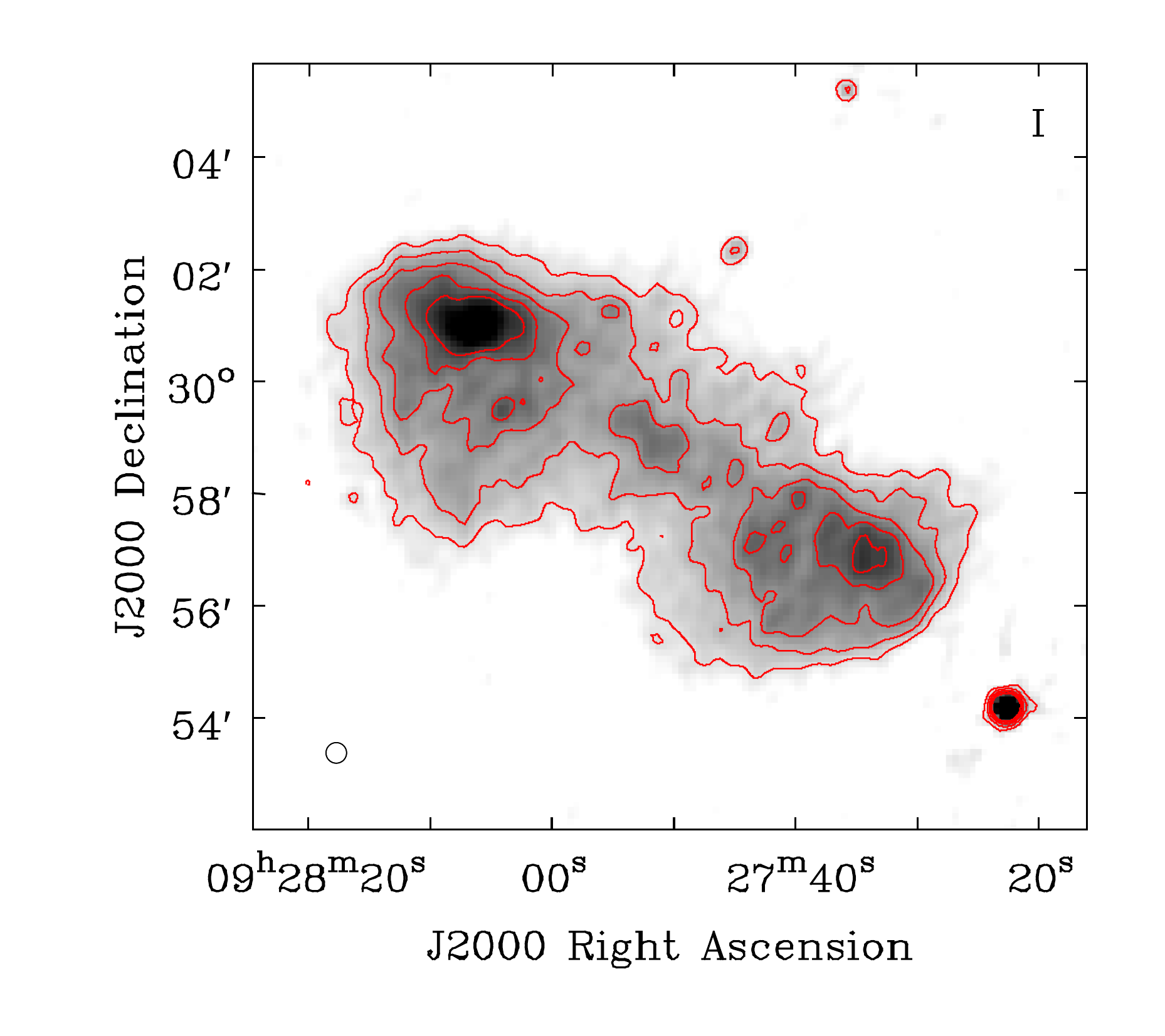}
\label{tgt_zoom:subfig1}}
\subfloat[][B2~0924+30 LOFAR HBA contours overlaid on an SDSSr image.]{\includegraphics[width=0.5\textwidth]{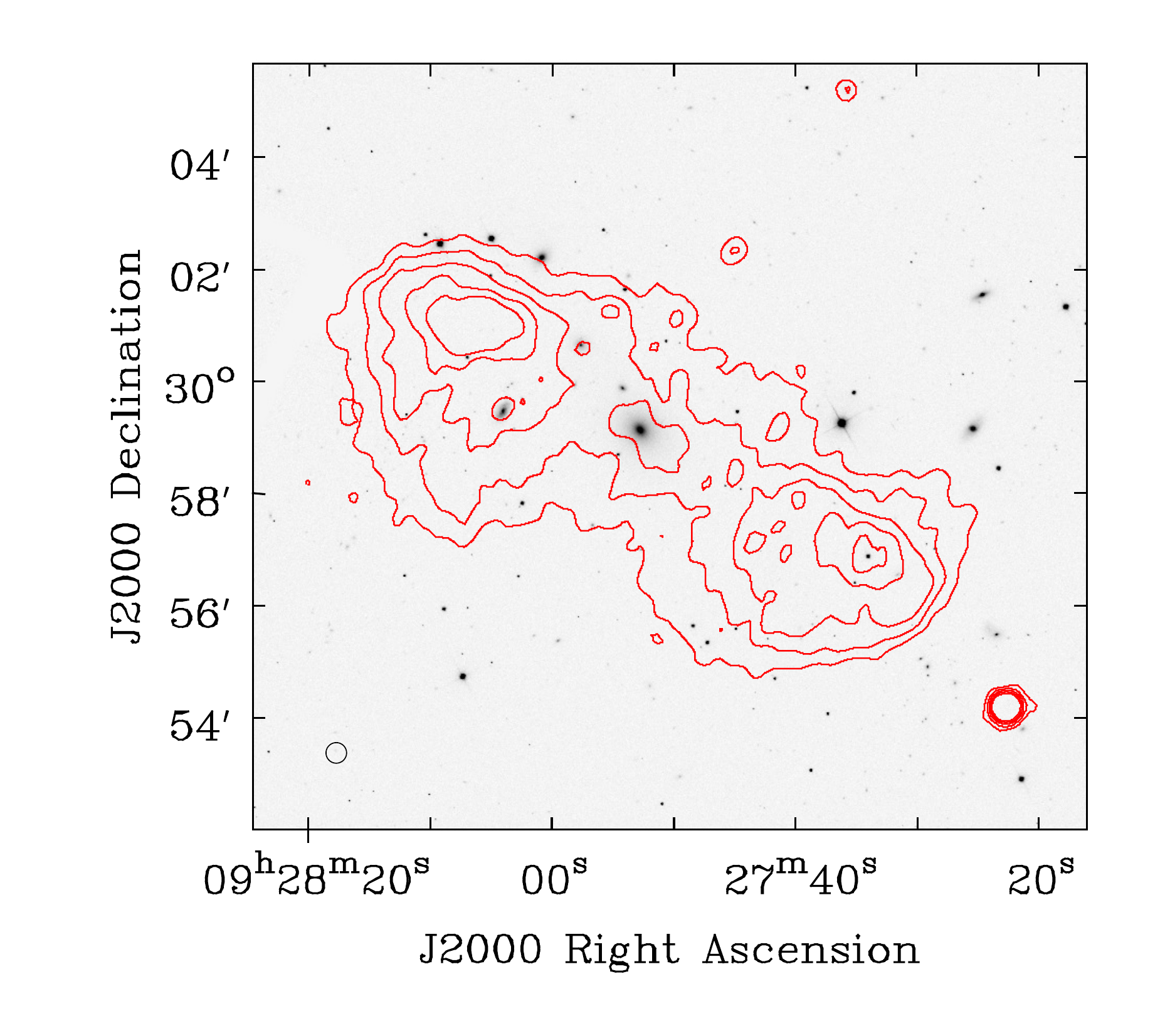}
\label{tgt_zoom:subfig2}}
\caption{B2~0924+30: LOFAR image obtained by averaging the higher resolution HBA images. Beam size: $ 22\arcsec\ \times 22\arcsec\ $, $ \sigma \, = \, 2 $ \mJybeam. Contour levels: $ (-3, 3, 6, 9, 12, 15) \cdot 2$ \mJybeam\ .}
\label{c5:tgt_zoom}
\end{figure*}

Figure \ref{c5:tgt_zoom} shows the LOFAR view of the target in the-high resolution (22\arcsec) averaged image (see Table \ref{c5:table:2}). We note increased surface brightness regions ($ > 30 $ \mJybeam) within the lobes that are located on opposite sides of the host galaxy. Also, there is an enhancement of surface brightness around the position of the host galaxy. The source is enveloped in a lower surface brightness cocoon.

Several smaller regions of increased surface brightness are noticeable within the radio lobes (see Fig. \ref{tgt_zoom:subfig2}). Two of them in the north-east (NE) lobe can be identified with background/foreground galaxies.

A point source located off the outermost edge of the SW lobe has been identified with a quasar \citep{RefWorks:16}.

There is no noticeable radio core at the position of the host galaxy as ascertained from our high-resolution imaging and images with 1\arcsec resolution obtained by \cite{RefWorks:97} at 5000 MHz. \cite{Giovannini1988} place an upper limit on the core flux density of $ S_{4900MHz} < 0.4 $mJy which, in relation to the total surface brightness, hints at the remnant nature of the radio source.

\subsection{Spectral analysis}
\label{c5:spec_an}

The morphology of B2~0924+30 supports its classification as AGN remnant, fading away after the AGN which has created it has shut down. Here, we elaborate on its spectral properties. The shape of the integrated flux spectrum encodes the activity history of a given radio source and can be a powerful tool in understanding the exact nature of the observed radio emission.

\subsubsection{Integrated spectrum}
\label{c5:int_spec_sec}

\cite{RefWorks:206} fitted a synchrotron ageing model to data collected from the literature as well as their own observations. We repeated the fitting procedure, adding the integrated flux density measured from our averaged LOFAR map. An overview of the measurements is given in Table \ref{c5:table:3}.

\begin{table}[!htpb]
\noindent \caption{\small B2~0924+30 flux density}
\label{c5:table:3}
\centering
\small
\begin{tabular}{ p{2.5cm} p{3cm} p{2cm}}
\hline\hline\\
\small $ \nu $ [MHz] & $ S_{\nu} $ [mJy] & Ref. \\
\hline\\
140 & $ 6306 \pm 1261 $ & 1\\
151 & $ 4600 \pm 360 $ & 2\\
325 & $ 2425 \pm 124 $ & 1, 5\\
609 & $ 1094 \pm 56 $ & 1, 3\\
1400 & $ 420 \pm 43 $ & 3\\
4750 & $ 60 \pm 7 $ & 3\\
10550 & $ 10 \pm 4 $ & 4\\
\hline
\end{tabular}
\tablebib{(1) this work; (2) \cite{RefWorks:97}; (3) \cite{RefWorks:206}; (4) \cite{RefWorks:246}; (5) \cite{RefWorks:138}}
\end{table}

The magnetic field strength was derived by assuming an equipartition between the energy contained in the magnetic field and in relativistic particles according to \cite{RefWorks:14}. In our calculations, we used a central frequency of 609 MHz, with a spectral index of $ \alpha \, = \, -1.2 $ (average over the source) and lobe extent of $ 4\farcm8 $. The cut-off frequency values for the calculation were taken to be $ 10 $ MHz and $ 10 \, 000 $ MHz, and the electron to proton ratio was set to unity. We computed the magnetic field value for each lobe separately and then averaged the result. Our estimate gives a value of $ 1.35 \, \mu $G \citep[similar to what is found for other remnant sources;][]{RefWorks:34, Brienza2016a}.

If we impose a low energy cut-off in the particle spectrum, instead of a low frequency cut-off in the emitted synchrotron radiation spectrum, for the magnetic field \citep{RefWorks:279} we get

\[ B \, = \, 1.18 \, \gamma_{\mathrm{min}}^{-0.3(3)} \, \left( B' \right)^{0.8(3)}, \addtag \]

\noindent where $ B' $ is the equipartition magnetic field that was calculated previously and $ \gamma_{\mathrm{min}} $ the low-energy cut-off value. For $ \gamma_{\mathrm{min}} \, = \, 1450 $, $ B \, = \, B' $, while for $ \gamma_{\mathrm{min}} \, = \, 500 $, $ B \, = \, 1.91 \, \mu $G (30 \% larger). Choosing an energy cut-off value is somewhat arbitrary. Based on equipartition arguments, \cite{RefWorks:206} derive a value of $ B \, = \,  1.6 \, \mu $G. We therefore decide to adopt the equipartition value that we initially derived and assume a magnetic field strength of $ B \, = \, 1.35 \, \mu $G for all subsequent analysis.

We fitted a continuous-injection model with an off phase \citep[KGJP,][]{RefWorks:188} to the integrated flux density measurements. Based on a modification of the expression found in \cite{Shulevski2015b}, the particle distribution function is

\[ N(t_{\mathrm{off}},t_{\mathrm{on}},b,\gamma,E) = \]
\[ \left\{ \begin{array}{ccc} \frac{E^{-(\gamma + 1)}}{b(\gamma - 1)((1 - bEt_{\mathrm{off}})^{\gamma - 1} - (1 - bE(t_{\mathrm{on}} + t_{\mathrm{off}}))^{\gamma - 1})} & \mbox{for} & E < \frac{1}{b(t_{\mathrm{on}} + t_{\mathrm{off}})} \\\\ \frac{E^{-(\gamma + 1)}}{b(\gamma - 1)(1 - bEt_{\mathrm{off}})^{\gamma - 1}} & \mbox{for} & \frac{1}{b(t_{\mathrm{on}} + t_{\mathrm{off}})} \leqslant E \leqslant \frac{1}{bt_{\mathrm{off}}} \\\\ 0 & \mbox{for} & E > \frac{1}{bt_{\mathrm{off}}} \end{array} \right., \addtag \]

\noindent where, $ t_{\mathrm{on}} $ and $ t_{\mathrm{off}} $ are the active phase duration and the time elapsed since source shut-down, $ b $ is a term describing the energy losses of the particles, and $ E \sim \sqrt{\nu/x} $ is the energy of the particles. $ x = \nu/\nu_{\mathrm{b}} $ represents the so-called scaled frequency. We assume a range of scaled frequencies, i.e. we do not fit for the break frequency explicitly. The energy-loss term was taken to be the one described by \cite{RefWorks:125}; hence the JP suffix in the model label

\[ b \sim B^{2}\left[\frac{2}{3} + \left(\frac{B_{\mathrm{IC}}}{B} \right)^{2}\right], \addtag \]

\noindent where $ B_{\mathrm{IC}} = \sqrt{\frac{2}{3}}B_{\mathrm{CMB}} $ is the effective inverse Compton magnetic field, and $ B_{\mathrm{CMB}} = 3.25(1 + z)^{2} $ is the equivalent cosmic microwave background (CMB) magnetic field.

The observed flux density is given by:

\[ S(\nu) = S_{0}\sqrt{\nu}\int F(x)x^{-1.5}N(x)dx, \addtag \]

\noindent where $ S_{0} $ represents a scaling factor, $ F(x) = x\int_{x}^{\infty} K_{5/3}(z)dz $ is defined by \cite{RefWorks:187} and $ K_{5/3} $ is the modified Bessel function.

The KGJP model is warranted since the integrated flux density includes contribution from formerly active source regions where particle acceleration was ongoing.

\begin{figure}[!]
\centering
\includegraphics[width=0.4\textwidth]{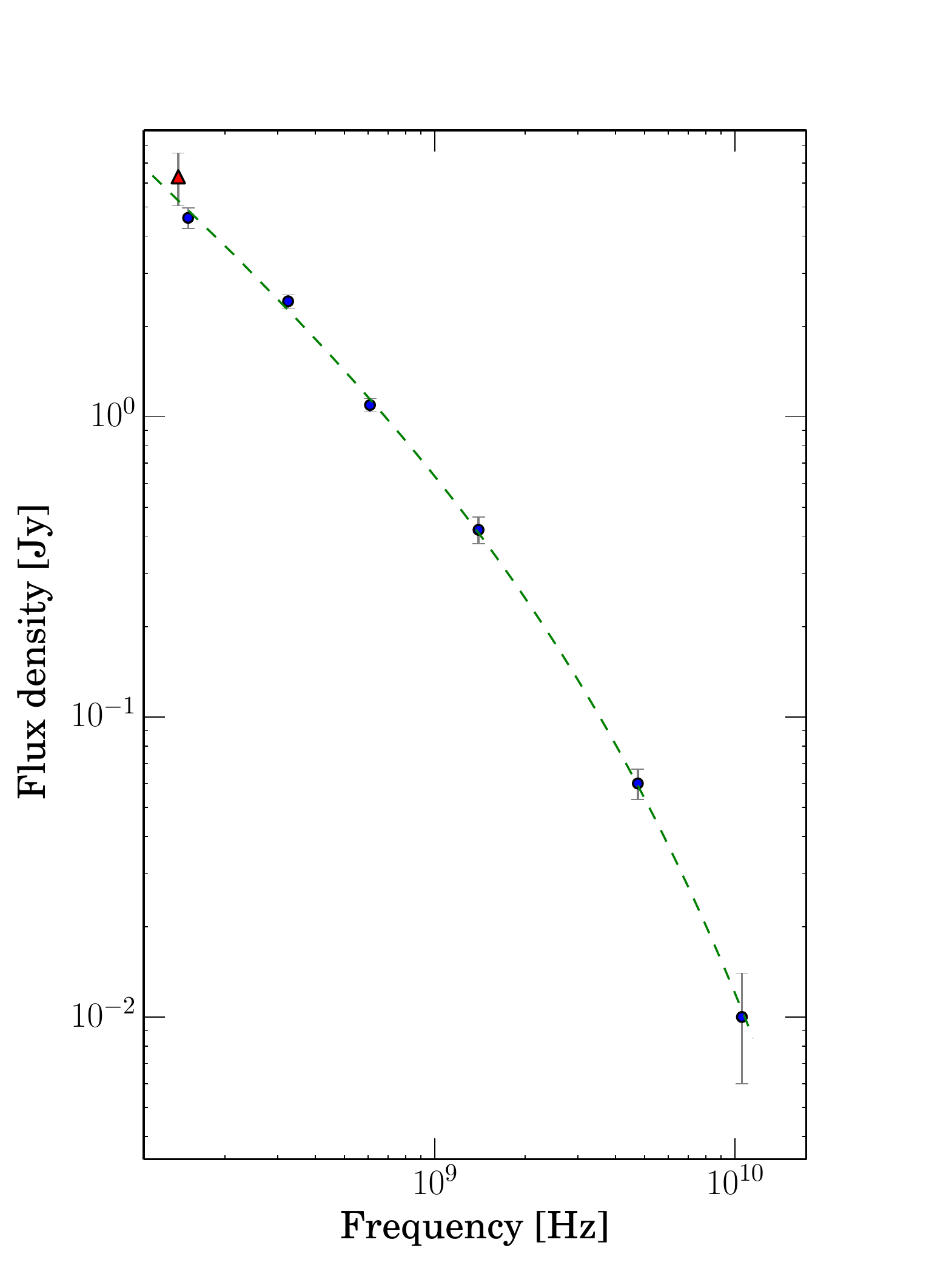}
\caption{Best fit Komissarov-Gubanov JP (KGJP) model to the integrated flux density measurements. The red triangle represents the LOFAR data point.}
\label{c5:int_spec}
\end{figure}

The best-fit values for the time during which the source was active and the time elapsed since the particle injection has ceased (time since shut-down) were found to be: $ t_{\mathrm{on}} \, = \, 55.65 \pm 2.25 $ Myr and $ t_{\mathrm{off}} \, = \, 32.04 \pm 1.57 $ Myr respectively. We assumed that the magnetic field is constant in time and over the source extent and we neglect adiabatic losses. The best-fit value for the injection spectral index (the spectral index of the particles immediately after they were accelerated / energized) was found to be $ \alpha_{\mathrm{inj}} \, = \, -0.85_{-0.1}^{+0.2} $. We used the Kapteyn package \citep{RefWorks:245} for the model-fitting. The model acceptance criteria are identical to those presented in \cite{Shulevski2015b}.
Our best fit-value for the total source age is $ t_{\mathrm{s}} = t_{\mathrm{on}} + t_{\mathrm{off}} \sim 88 $ Myr. Our derived ages differ from those reported by \cite{RefWorks:206} of $ 54_{-11}^{+12} $ Myr (they also assume a constant magnetic field strength and neglect adiabatic losses). We are, however, in agreement with their derived value for the injection spectral index ($ \alpha_{\mathrm{inj}} \, = \, -0.87 \pm 0.09 $). The values we derive for the epochs of source activity are not directly comparable to those of \cite{RefWorks:206} since they used an ageing-only (JP) model in their integrated flux density spectral analysis. The best model fit is shown in Fig. \ref{c5:int_spec}. While the spectral curvature is expected for a remnant radio source, the steepness of the injection spectral index (confirmed by our LOFAR measurements) is puzzling.

\subsubsection{Spectral index and curvature maps}
\label{c5:spec_curv_maps}

To study the plasma properties in the remnant lobes, we have produced the highest resolution spectral index map of B2~0924+30 at low frequencies to date. We averaged together the lower resolution HBA images to a single low frequency image (140 MHz) and used the 609 MHz WSRT image from \cite{RefWorks:206}. The data sets have a closely matching UV coverage. The spectral analysis input images were smoothed to a resolution of 60\arcsec and registered to the same pixel size. We derived the spectral index in the standard manner, and propagated the errors of the flux-density to get an estimate of the error in determining the spectral index. We assumed that the flux-density errors in both maps are uncorrelated. We did this for each pixel above a $ 7\sigma $ level in the input images.

\begin{figure*}[!htpb]
\centering
\subfloat[][$ \alpha_{140}^{609} $ spectral index map]{\includegraphics[width=0.5\textwidth]{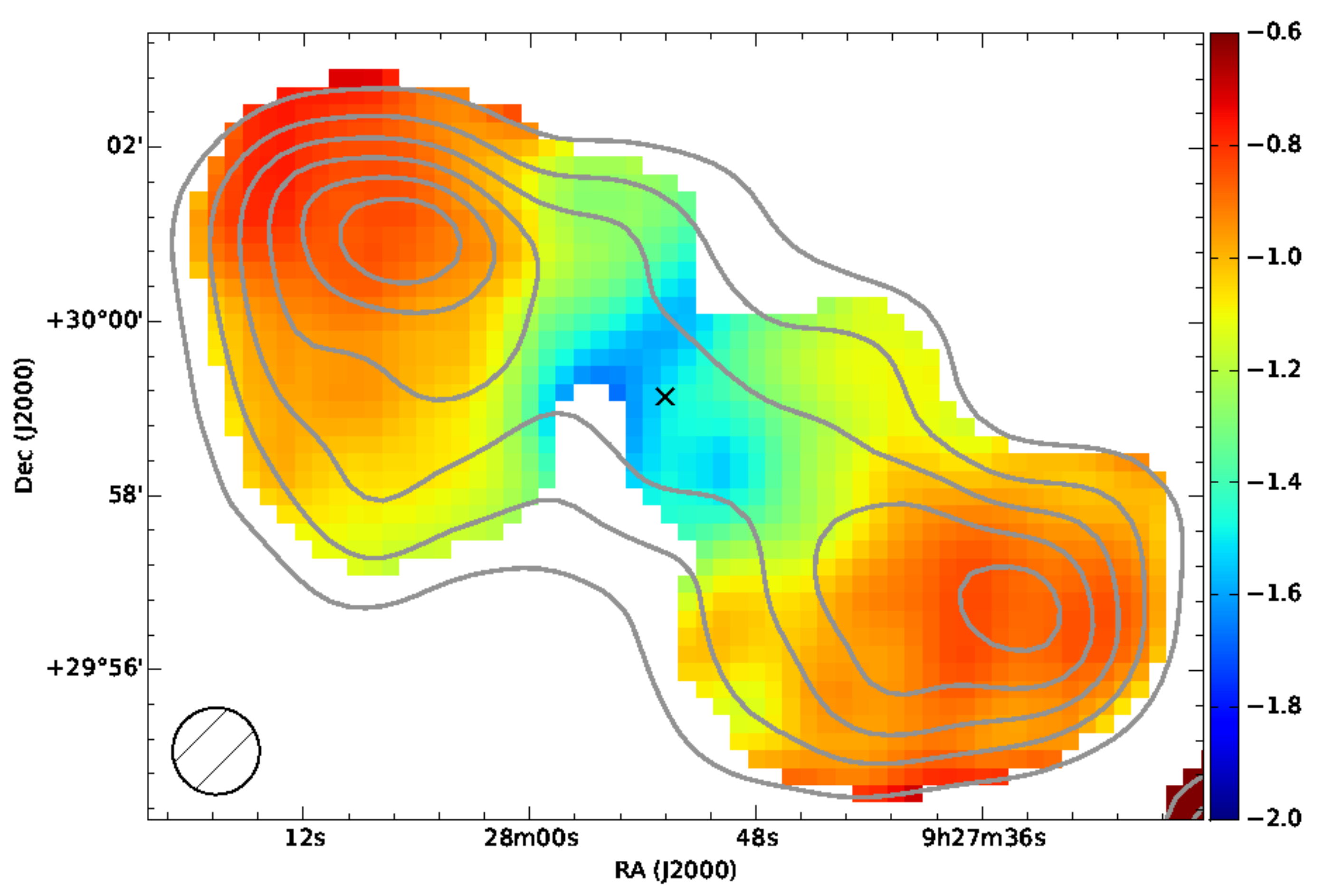}
\label{c5:spec_maps:subfig1}}
\subfloat[][Spectral index error map]{\includegraphics[width=0.5\textwidth]{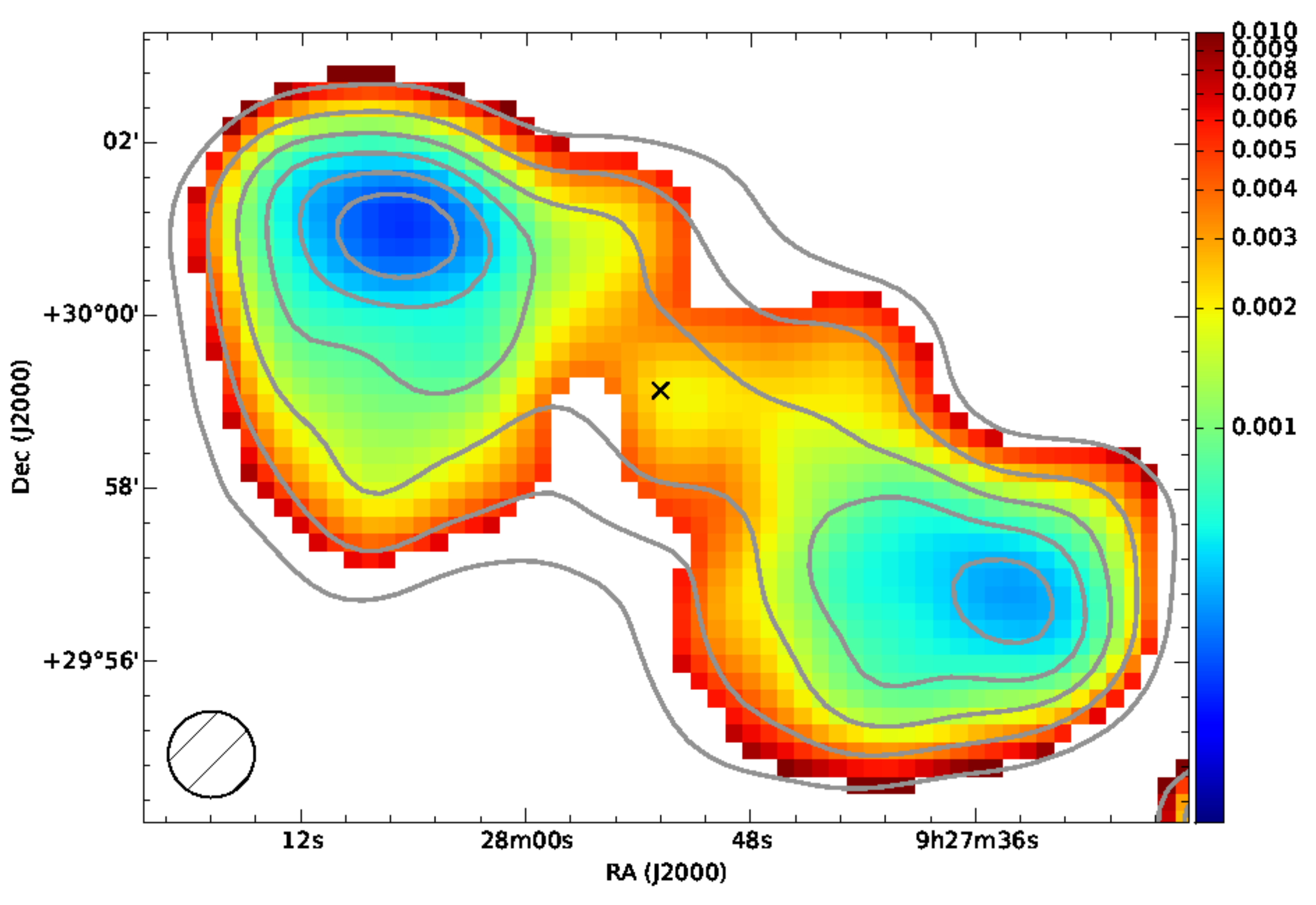}
\label{c5:spec_maps:subfig2}}
\quad
\subfloat[][$ \alpha_{140}^{609} \, - \, \alpha_{609}^{1400} $ spectral curvature map]
{\includegraphics[width=0.5\textwidth]{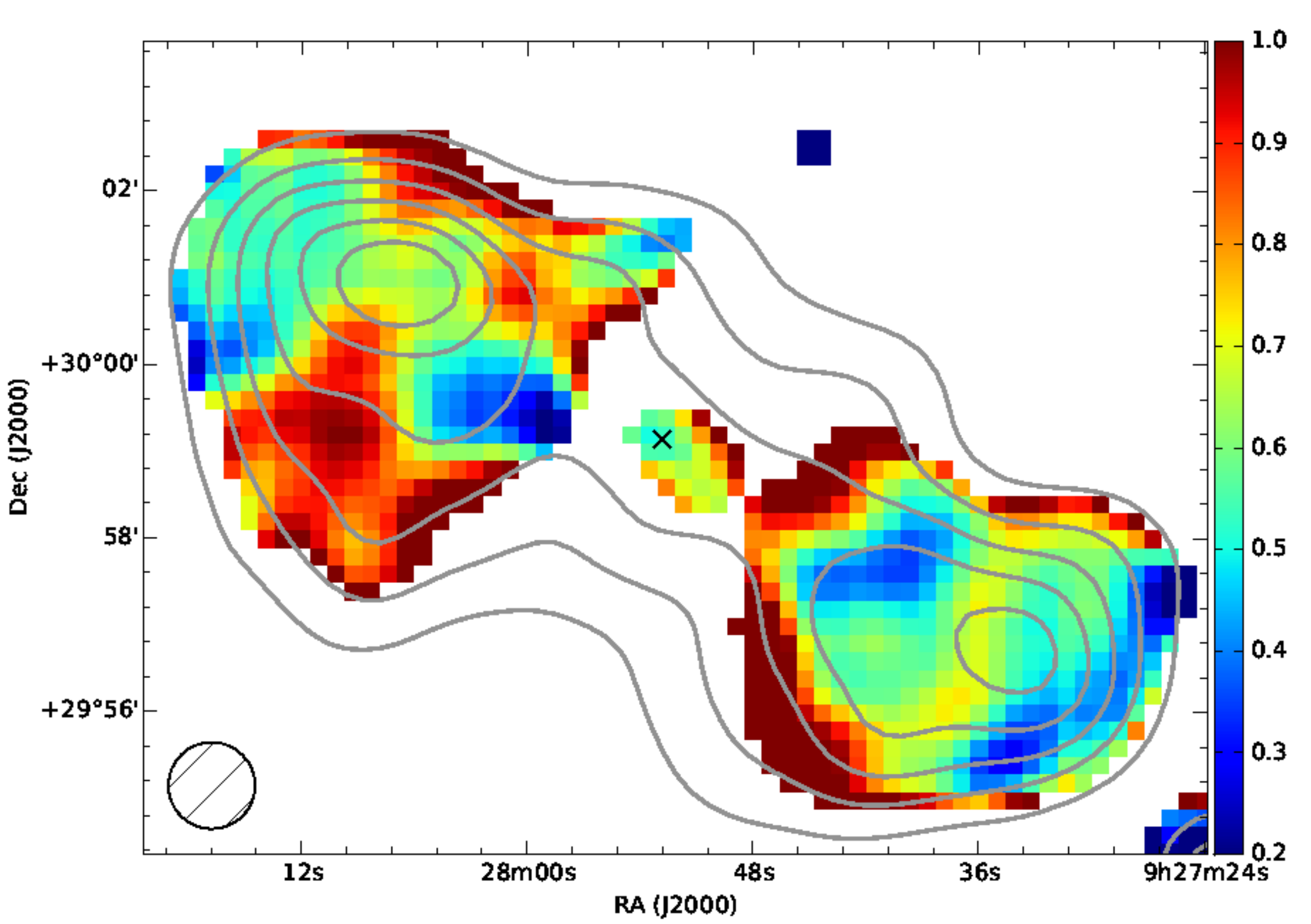}
\label{c5:spec_maps:subfig3}}
\subfloat[][Spectral curvature error map]{\includegraphics[width=0.52\textwidth]{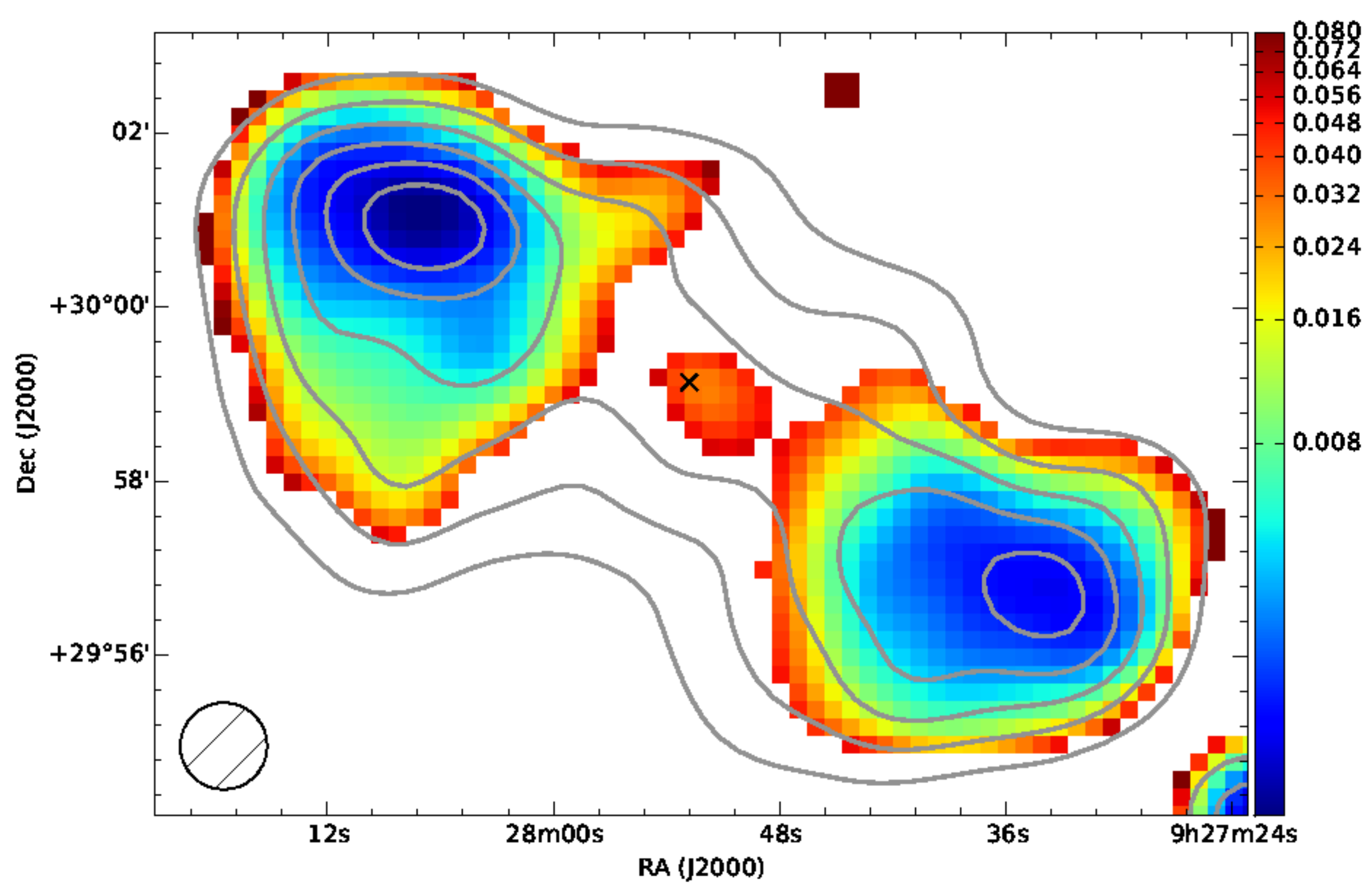}
\label{c5:spec_maps:subfig4}}
\caption{Spectral index and spectral curvature maps for pixels with surface brightness greater than $ 7\sigma $ and $ 3\sigma $, respectively, in all of the input maps. We used the averaged low-resolution LOFAR image (Table \ref{c5:table:2}). Overlaid are LOFAR contour levels spanning the interval between $ -10 \sigma $ and $ 60\sigma $, with a step of $ 10 \sigma $, where $ \sigma \, = \, 4 $ \mJybeam. The black cross indicates the position of the host galaxy.}
\label{c5:spec_maps}
\end{figure*}

In Figs. \ref{c5:spec_maps:subfig1} and \ref{c5:spec_maps:subfig2}, we can see that the spectral index in the lobes varies from $ \alpha \, \sim \, -1.4 $ at their inner edges, to $ \alpha \, \leq \, -0.75 $ at the outer edges. The average value for the integrated spectral index of the source is relatively steep at low frequencies, around $ \alpha \, \sim \, -1 $, in agreement with previous studies \citep{RefWorks:206}, as well as with the injection spectral index we obtained previously from fitting the integrated spectrum in Sect. \ref{c5:int_spec_sec}.

We observe that the lobes have steep spectral index values that flatten out going towards the outermost lobe edges; this is especially prominent in the NE lobe.

We also derived a spectral curvature map ($ \mathrm{SPC} \, = \, \alpha_{140}^{609} \, - \, \alpha_{609}^{1400} $) in an analogous fashion to the spectral index map, using an NVSS survey\footnote{The NVSS image (Table \ref{c5:table:2}) is missing flux on large angular scales. The integrated flux-density value at 1400 MHz that we use (Table \ref{c5:table:3}) is taken from \cite{RefWorks:206}, who used single dish Effelsberg telescope measurements to correct for the loss. Based on inspection of the corrected image \citep[Fig. 5a in][]{RefWorks:206}, we conclude that our spectral curvature and ageing derivation (for the lobe and core regions) is not affected by us using the uncorrected NVSS map.} image of the target as the highest frequency data point. We derived the spectral curvature for pixels above a $ 3\sigma $ level in all of the input images, to be able to map the regions around the host galaxy. The results are shown in Figs. \ref{c5:spec_maps:subfig3} and \ref{c5:spec_maps:subfig4}.

In line with our previous discussion, the curvature map provides interesting insights into the spectral properties of the source. The remnant lobes reveal more structure, with some areas showing large curvature up to $ \mathrm{SPC} = 1 $. This suggests that different regions have spectral breaks at different frequencies, which indicates different radiative ages. For example, the lateral lobe edges show pronounced spectral steepening at higher frequencies.

\subsection{Radiative ages}
\label{c5:rad_ages}

\begin{figure*}[!htpb]
\centering
\subfloat[][Radiative age]{\includegraphics[width=0.5\textwidth]{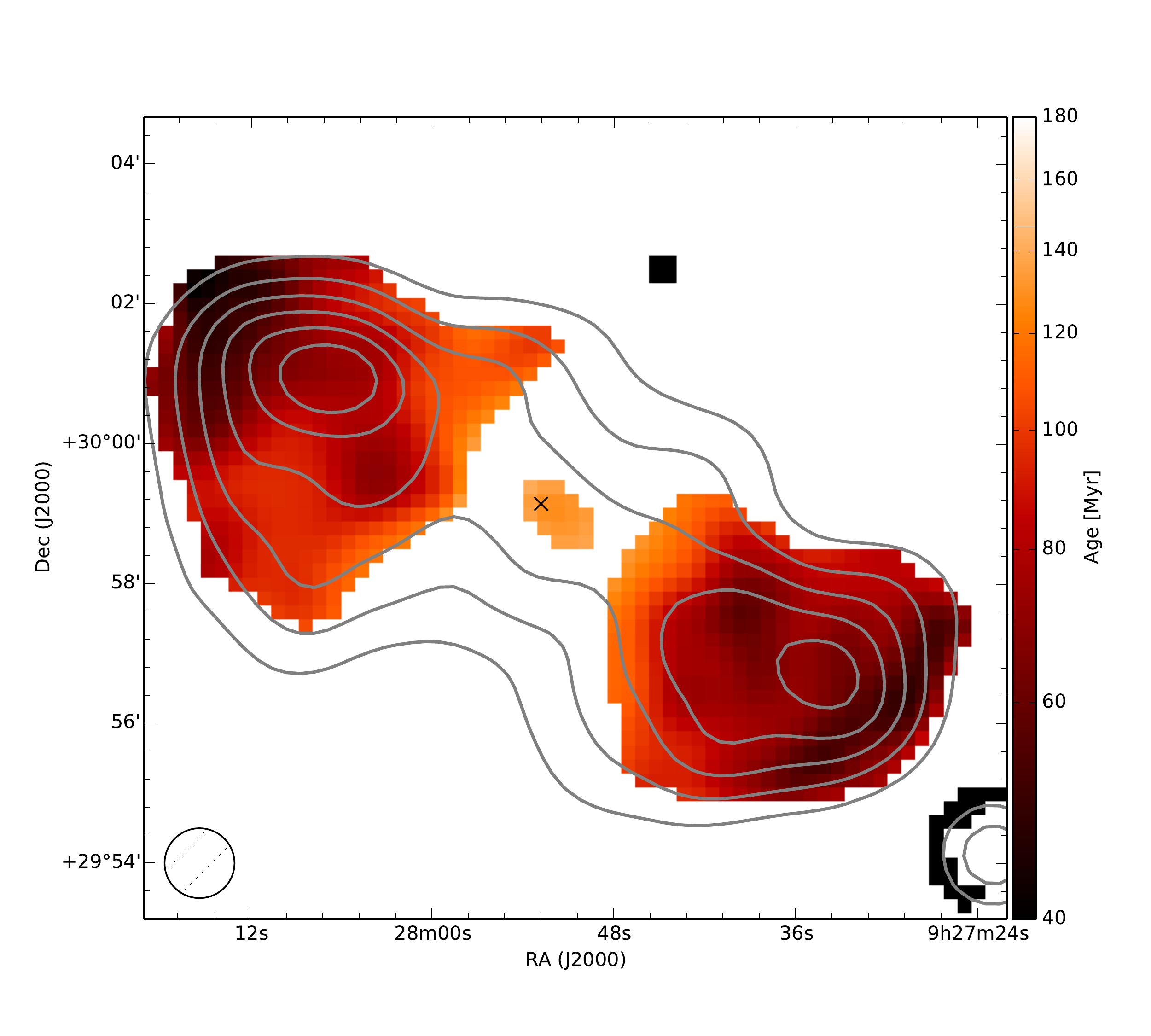}
\label{c5:age_maps:subfig1}}
\subfloat[][Age error]{\includegraphics[width=0.5\textwidth]{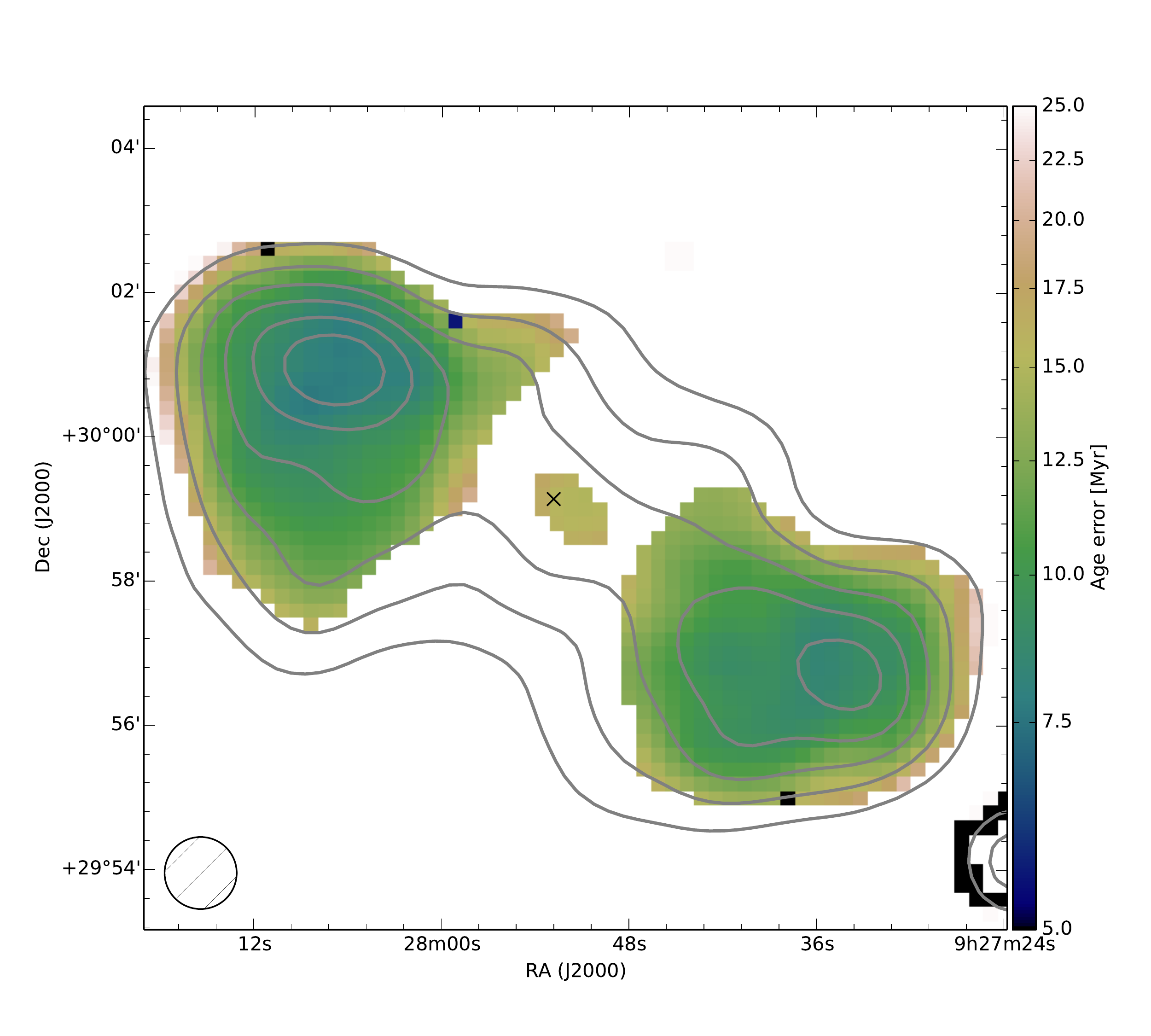}
\label{c5:age_maps:subfig2}}
\caption{Radiative ages and age errors derived from fitting an ageing-only model using a JP-loss term to the data for $ \alpha_{inj} = -0.85 $. Contours are the same as in Fig. \ref{c5:spec_maps}.}
\label{c5:age_maps}
\end{figure*}

To gain a better insight into the activity history of the radio source, we took our averaged LOFAR image, together with an 609 MHz WSRT image and an NVSS survey map (Table \ref{c5:table:2}), and fitted a JP ageing-only model with a particle distribution function:

\[ N(t_{\mathrm{off}},b,\gamma,E) = \]
\[ \left\{ \begin{array}{ccc} E^{-\gamma}(1 - bEt_{\mathrm{off}})^{\gamma - 2} & \mbox{for} & E \leqslant \frac{1}{bt_{\mathrm{off}}} \\\\
0 & \mbox{for} & E > \frac{1}{bt_{\mathrm{off}}} \end{array} \right., \addtag\]

\noindent to produce an age map for the source, shown in Fig. \ref{c5:age_maps}.
The injection index was not fitted for; its value was fixed to the one found ($ \alpha_{\mathrm{inj}} \, = \, -0.85 $) during the integrated flux density spectrum-fitting in Sect. \ref{c5:int_spec}. The magnetic field strength used was also the same as we used earlier, $ B \, = \, 1.35 \mu$G.

Age mapping provides more information compared to age-model fitting to integrated flux-density measurements for a given source. In the case of B2~0924+30, for the youngest regions at the edges of the lobes we determined an age of around 50 Myr; the source age increases as we look toward the host galaxy; the lobe inner edges show ages of up to 120 Myr, and the center regions around 150 Myr. These findings are in agreement with the ageing profile reported by \cite{RefWorks:206}. Since we fitted an ageing-only model (JP), we have an estimate for the time elapsed since the plasma was last energized across the source.

Consequently, for resolved sources, we can estimate the duration of their active phase as the difference between the oldest and youngest age read-off from the map: $ t_{\mathrm{on}} = t_{\mathrm{max}} - t_{\mathrm{min}} $. For B2~0924+30, we find an active phase duration of around 100 Myr. Furthermore, the elapsed time since the shut-down is given by the youngest age found ($ t_{\mathrm{off}} = t_{\mathrm{min}} $) and, in the case of B2~0924+30, this is found to be around 50 Myr.

The age mapping was done with a more limited spectral coverage than the integrated spectral index fit which we performed in Sect. \ref{c5:int_spec_sec}. The reason is that the available maps at frequencies higher than 1400 MHz were of lower resolution (> 1\arcmin) and less sensitive to extended emission. Even so, the mapping shows that (as expected) the source age derived from a (KGJP) model fit to the integrated spectrum is only an estimate for the total source age.

The source age we derived using age mapping is around two times higher than the age obtained by model fitting to the integrated flux-density measurements reported in Sect. \ref{c5:int_spec_sec}.

The ages derived for the inner lobe regions of B2~0924+30 are comparable to the oldest sources in, for example, the sample of \cite{RefWorks:34}.

\begin{figure}[!htpb]
\centering
{\includegraphics[width=0.5\textwidth]{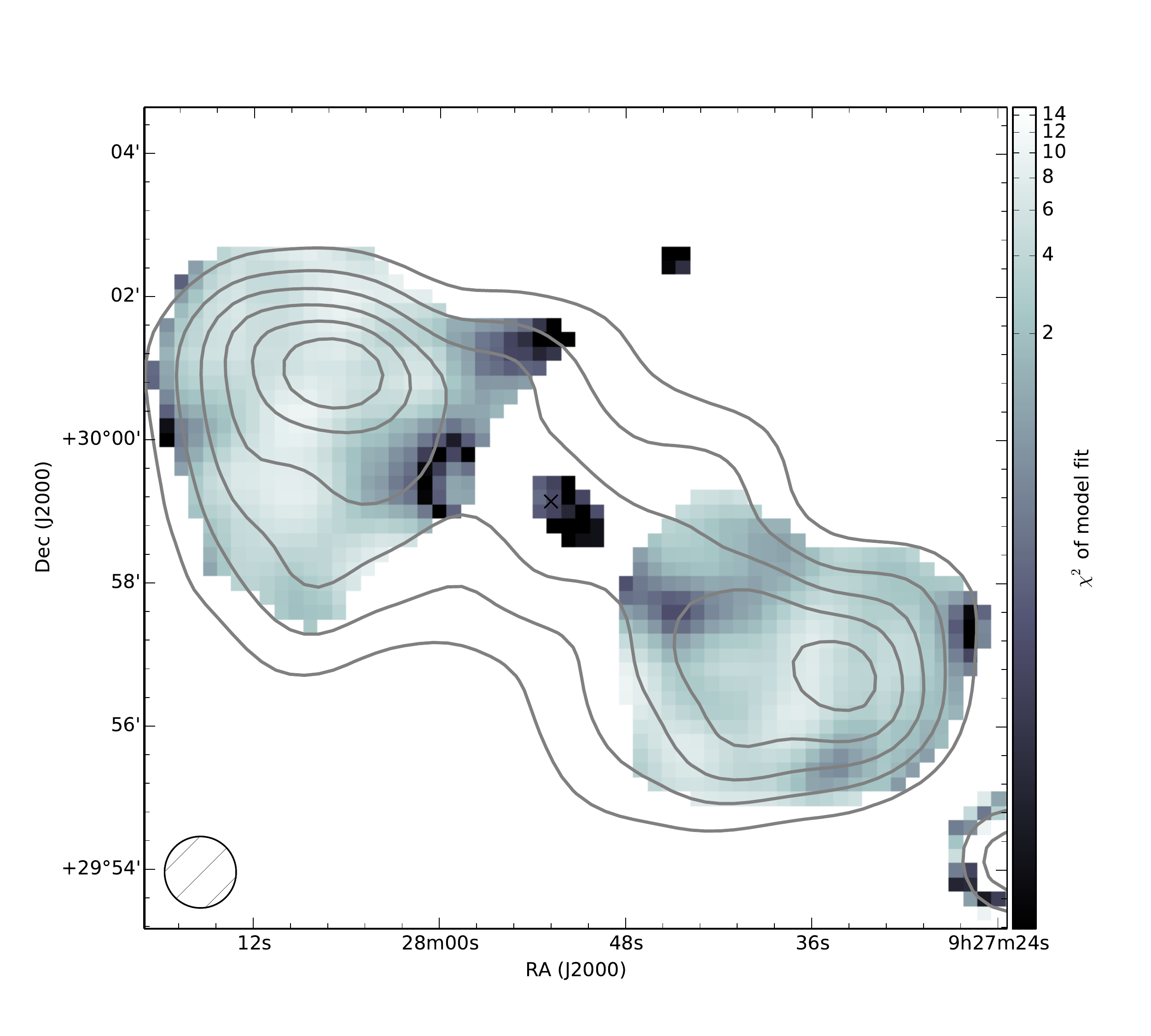}}
\caption{$ \chi^{2} $ values of the model fit (measuring the goodness of the fit for one degree of freedom), as defined in \cite{Shulevski2015b}. Contours are the same as in Fig. \ref{c5:spec_maps}.}
\label{c5:fit-err}
\end{figure}

\subsection{Spectral shifts}
\label{c5:spec_shift}

Investigating the source energetics can be done in an analogous manner to the spectral curvature map (Fig. \ref{c5:spec_maps:subfig3}) by plotting the low- and high-frequency spectra in a "colour-colour" plot \citep{RefWorks:189}. This approach enables us to visualize the spectral shapes of different source regions and compare them with (spectral) ageing models.

\begin{figure*}[!htpb]
\captionsetup[subfigure]{labelformat=empty}
\centering
\subfloat[][]{\includegraphics[width=0.5\textwidth]{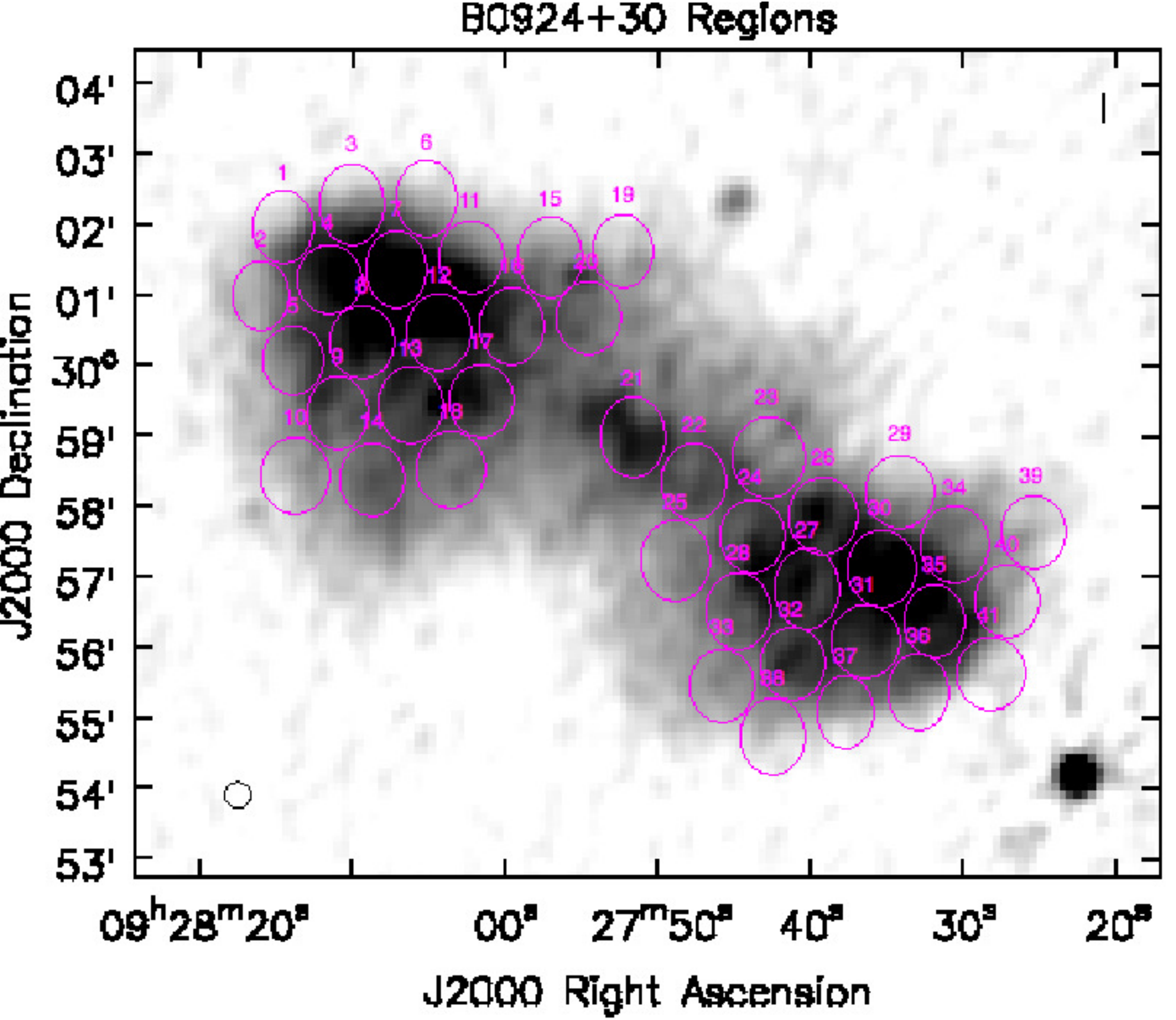}}%
\subfloat[][]{\includegraphics[width=0.4\textwidth]{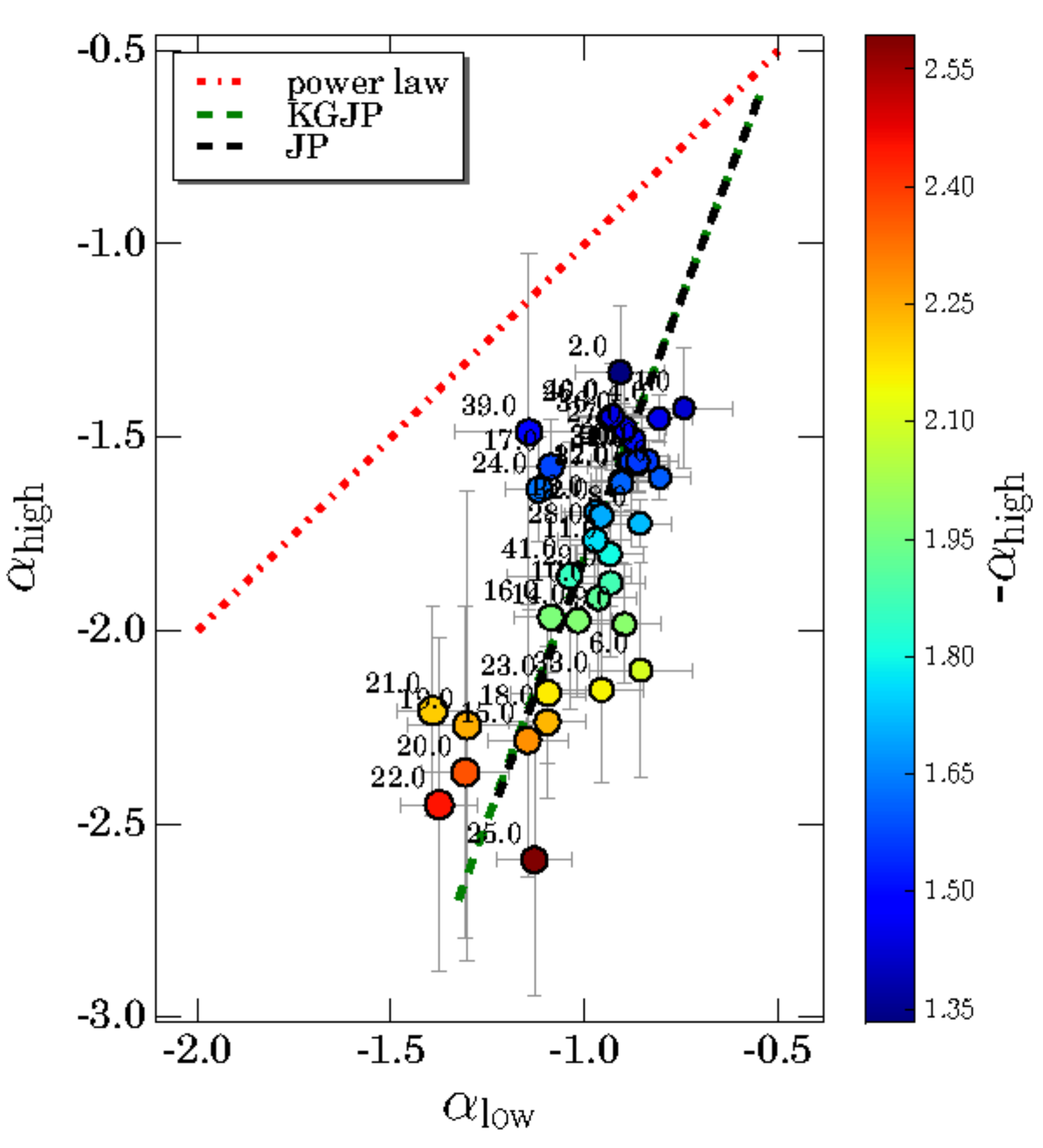}}%
\quad
\centering%
\subfloat[][]{\includegraphics[width=0.48\textwidth]{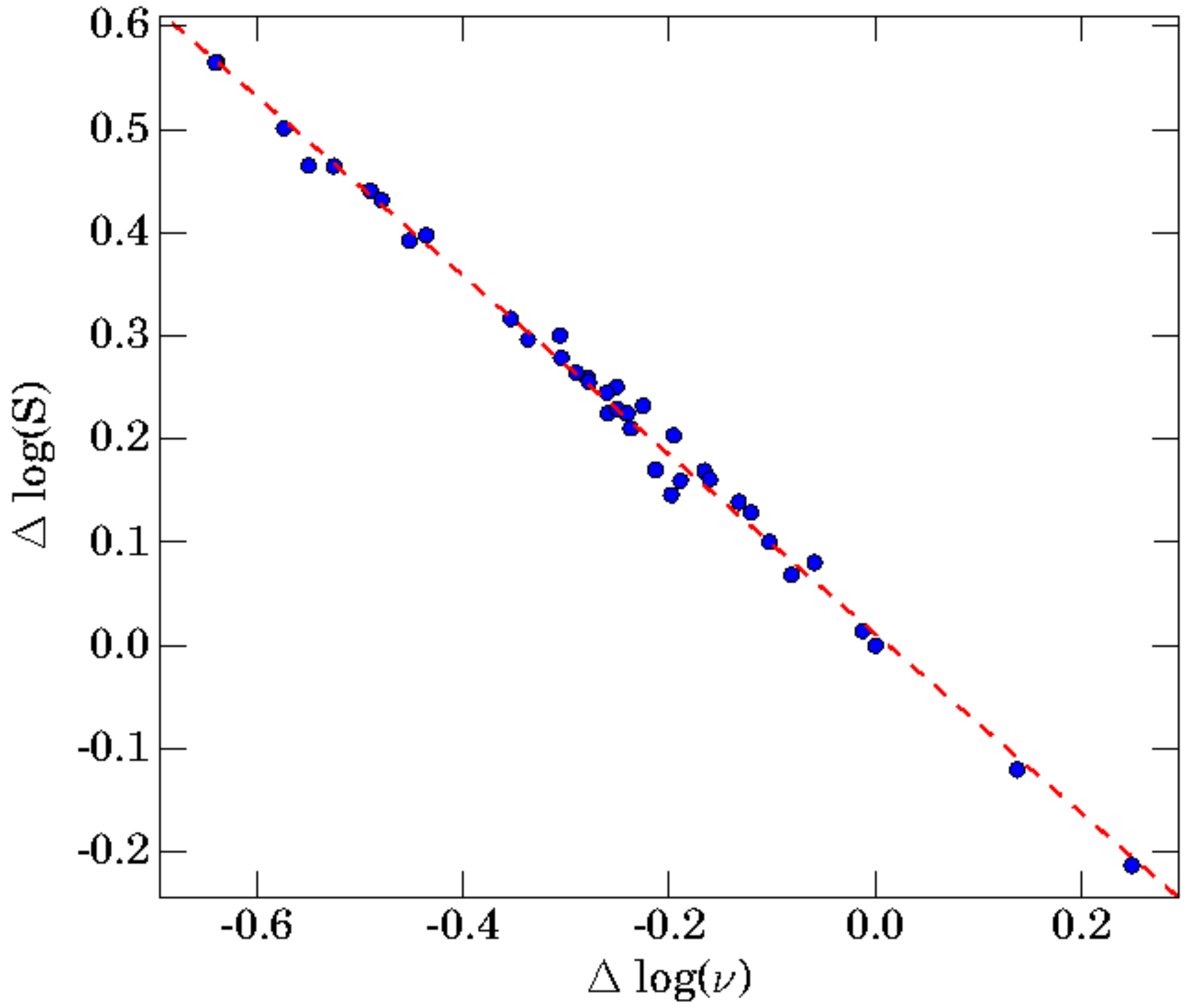}}%
\subfloat[][]{\includegraphics[width=0.42\textwidth]{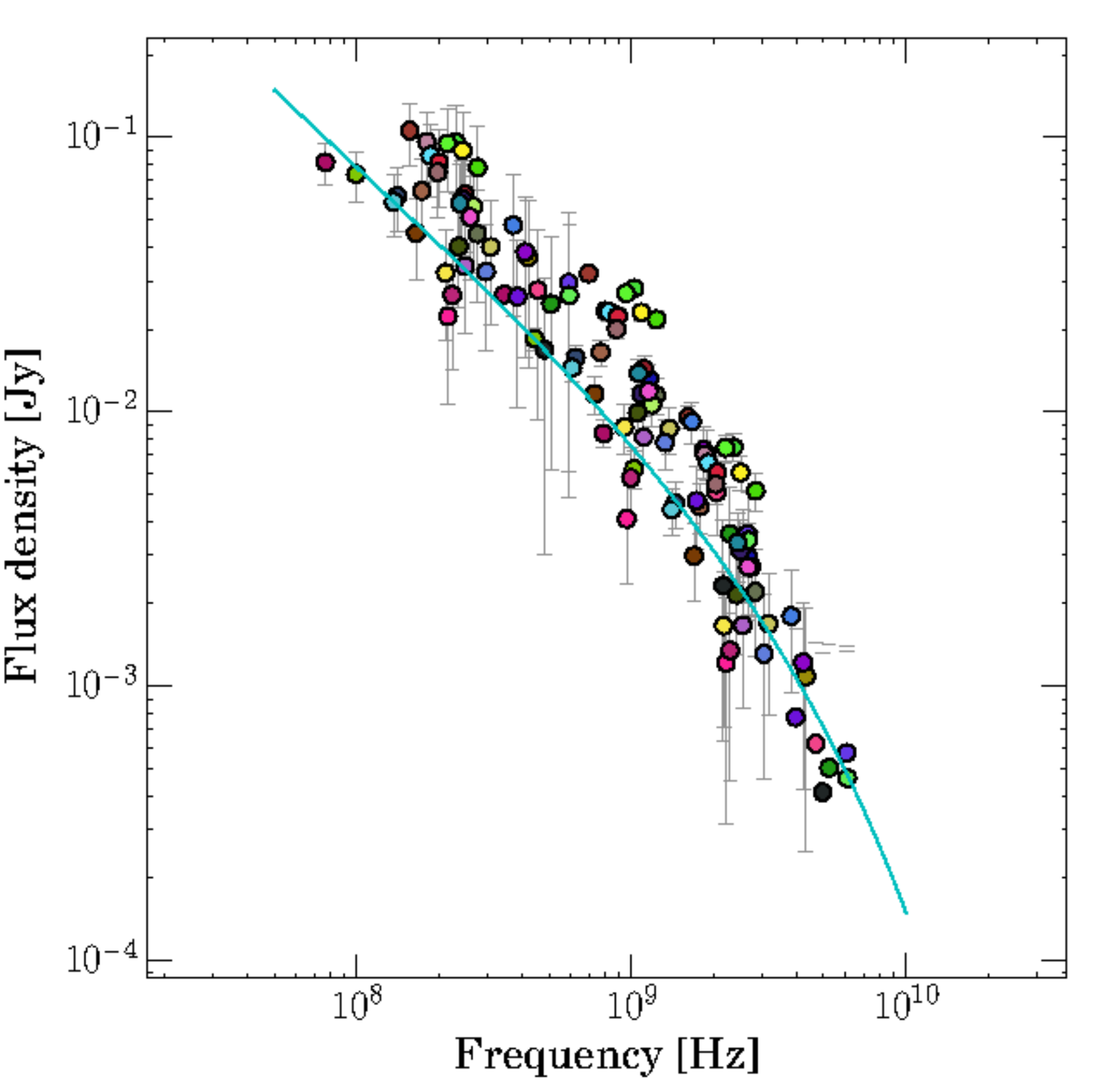}}%
\caption{\textbf{Top left: }Measurement regions overlaid on a LOFAR gray-scale map of the source. \textbf{Top right: }Colour-colour plot. The colour of the plotted points indicates their high spectral index value, while their size is proportional to their low spectral index value. Here, $ \alpha_{\mathrm{low}} = \alpha_{140}^{609} $ and $ \alpha_{\mathrm{high}} = \alpha_{609}^{1400} $ \textbf{Bottom left: }Shift plot. The slope of the fit is $ -0.87 \pm 0.01 $ \textbf{Bottom right: }Total set of shifted data and a (JP) model fit for all of the regions. Data points belonging to a given region share the same colour.}
\label{c5:shifts}
\end{figure*}

To this end, we performed the analysis on the same data set we used in the ageing modelling, described above. Forty-one measurement regions were used (shown in Fig. \ref{c5:shifts}, top left panel); their spectra were plotted on the $ \alpha_{140}^{609} $ - $ \alpha_{609}^{1400} $ colour-colour plane. In the same plot, we show the loci of points occupied by different ageing models, as well as a simple power law. The distribution of points that represent the regions shows that the spectral shape, for most of them, is best fitted by synchrotron radiation from an aged plasma (Fig. \ref{c5:shifts}, top right panel).

We can reconstruct the global spectrum of the source by shifting the spectra (in the $ log(S) - log(\nu) $ plane) of each region so that, finally, all of the spectra line up. We perform the shifting by choosing a reference region and aligning the spectral breaks of the remaining regions with the break of the chosen region. The spectral breaks were computed by following the expression given by \cite{RefWorks:126}:

\[ \nu_{\mathrm{b}} \, \approx \, 3.4 \times 10^{5} B^{-3} t_{\mathrm{off}}^{-2}, \addtag \]

\noindent where $ \nu $ is expressed in GHz, $ B $ in $ \mu $G, and $ t $ in Myr. This technique has been implemented in a number of cases, for example \cite{RefWorks:193}, and it has the advantage of extending the spectral analysis over a wider range in frequency. If there is a global electron energy distribution across the source, the spectra of the individual regions trace it, and the differences between them are due to energy losses (radiative, adiabatic), magnetic field variations, and variations in electron density. By shifting the spectra, we account for these effects. The shift results are given in Fig. \ref{c5:shifts} (bottom right panel).

The amount of shift needed in the flux density-frequency plane per region is shown in Fig. \ref{c5:shifts} (bottom left panel). We see that the shift values lie on a straight line; this is a strong indicator that the spectral shape of the regions is dominated by radiative losses. If we sample regions with similar physical conditions, the slope of the fit represents the injection spectral index. In our case, the slope value is $ -0.85 $, which is in line with the best-fit injection spectral index that we arrived at in Sect. \ref{c5:int_spec_sec}, which provides inner consistency to our analysis.

\section{Discussion}
\label{c5:disc}

B2~0924+30 is an AGN radio remnant, a leftover from the time when it was an active radio galaxy. The remnant radio lobes are very well outlined, which may indicate confinement by the IGM. Our analysis shows that the youngest plasma is located at the outer lobe edges. Regions closer to the host galaxy are progressively older, and the diffuse radio emission at the position of the host galaxy (noticeable in the LOFAR image) is the oldest region of the source. There is no sign of restarted AGN activity.

In Sect. \ref{c5:res} we performed a detailed spectral index, curvature, and ageing analysis, mapping these quantities with the highest spatial resolution to date (extending to low frequencies) and comparing our results to previous studies. We found that the values derived for the spectral age using integrated flux density measurements are higher than values obtained in previous studies. The injection spectral-index values are in agreement with the literature and point to a steeper ($ \alpha_{\mathrm{inj}} \sim -0.85 $) injection index than is usually assumed.

Comparing our low frequency spectral index map (Fig. \ref{c5:spec_maps:subfig1}) with studies performed using higher frequency data \citep{RefWorks:206}, we infer that the lobe spectral index is somewhat flatter; this is most pronounced at the outer edges of the lobes. At the frequencies we study with LOFAR, we would expect that the spectrum is a power law that is representative of the injection spectral index. Recent studies \citep{Harwood_2017b} have shown that integrated flux-density models (incorporating a continuous injection phase) are unreliable in the recovery of parameters. They tend to depend on the frequency coverage and may provide systematically higher values for the injection spectral index. Further, determining the injection spectral index from the low-frequency spectrum alone is problematic. \cite{Harwood_2017a} show that the hotspot spectrum in maps of active sources is affected by stochastic acceleration and / or absorption processes that flatten the spectrum at low frequencies. At present, it is not known how (or whether) these processes reflect on remnant hotspot regions in inactive radio galaxies. Hence, it is uncertain how accurately we can determine the injection spectral index. In any case, within the stated errors, the injection index is in approximate agreement with studies of active radio galaxies \citep{Harwood_2015, Harwood2016}. Furthermore, simulations suggest \citep{Kapinska} that the observed integrated spectrum can steepen owing to mixing of electron populations.

The appearance of the core regions with their steepest spectral index, as well as the spectral index and age gradient, suggest that B2~0924+30 is a fading FRII source; the youngest regions are found towards the outer edges of the lobes, and the oldest are the regions towards the host galaxy. Taken together, these inferences point to the fact that we are observing source regions with particles that were last energized just before source shutdown. Our mapping suggests that we are seeing the remnants of hotspots which are prominent features in active FRII radio galaxies.

We were able to estimate a limit to the duty cycle for B2~0924+30 using the duration of the active and dormant periods that were read off the derived age maps. The time elapsed since it has shut down is estimated to be half of that spent in an active phase. This is in line with the DDRG sources studied by \cite{RefWorks:178} that we mentioned earlier (which, morphologically, appear to be restarted FRII sources) and different from the case presented in \cite{Brienza2016a}, which shows a significantly shorter active compared to dormant phase and seems to have FRI morphology ($ L_{\mathrm{1400MHz}} \, = \, 1.5 \times 10^{24} $\whz).

The spectral index and spectral ageing maps we produced not only support the claim that this is an FRII AGN remnant; the model fitting produces plasma ages that are higher than those derived using integrated flux density data.

The discrepancies between the ages derived from the integrated flux density analysis and resolved studies are model dependent. The integrated flux density analysis averages over source regions with different physical properties and consequently particle activity histories. Thus, the ages derived from the integrated flux density analysis may be different to that found using age mapping.

In Sect. \ref{c5:spec_shift}, we show that the dominant energy-loss mechanisms are inverse Compton and synchrotron radiation, suggesting that adiabatic expansion energy losses are negligible. This hints at the possibility that the radio plasma is somewhat confined. The presence of a low significance, possibly extended ROSAT X-ray emission detection that is associated with the host galaxy \citep{Canosa1999} may be relevant in this regard.

The fact that we can detect remnant hotspots suggests that relatively short time has elapsed since source shut-down (estimated at $ 50 $ Myr).

The paucity of AGN radio remnants may be due to long AGN duty cycles (the time elapsed between active phases being longer than the radio plasma lifetime) and/or the radio plasma ageing more rapidly owing to expansion losses.

A larger sample of AGN radio remnants is needed to put firmer constraints on these assumptions.

\section{Conclusions}
\label{c5:fin}

We have used LOFAR to obtain images of B2~0924+39 at low frequencies with the highest spatial resolution yet obtained for this source. This has enabled us to produce detailed spectral index maps and derive radiative ages over the extent of the source. We confirmed previous inferences \citep{RefWorks:206} that are consistent with this source being a FRII remnant. We have also shown that there is a continuum of ages that increase from the outer lobes to the regions at the position of the host galaxy.

In addition, we have demonstrated the detection of remnant hotspot regions at the outer lobe edges, further supporting the FRII nature of this source. This result highlights the value of high-resolution, high surface-brightness sensitivity LOFAR maps obtained at low frequencies in studying remnant radio galaxies, disentangling their nature and activity history.

We have shown that age estimates obtained from integrated flux density measurements differ from those obtained using resolved studies. This finding indicates caution; ages obtained by fitting models to integrated flux-density measurements tend to be affected by the activity history of those sources, or only provide limits to the ages obtained by mapping. 

Detailed studies of larger samples of AGN remnants are needed to answer the question as to whether the derived timescales are typical and consistent for this source type, and infer the details of their duty cycles.

\begin{acknowledgements}

The authors thank the anonymous referee for the useful comments and suggestions that helped improve this paper.\\
LOFAR, the Low Frequency Array designed and constructed by ASTRON, has facilities in several countries that are owned by various parties (each with their own funding sources), and that are collectively operated by the International LOFAR Telescope (ILT) foundation under a joint scientific policy.\\
RM gratefully acknowledges support from the European Research Council under the European Union's Seventh Framework Programme (FP/2007-2013) / ERC Advanced Grant RADIOLIFE-320745.\\
GJW gratefully acknowledges support from The Leverhulme Trust.\\
This work has made use of the NASA/IPAC Extragalactic Database (NED), which is operated by the Jet Propulsion Laboratory, California Institute of Technology, under contract with the National Aeronautics and Space Administration.\\
This work has made use of python (www.python.org), including the packages numpy (www.numpy.org), scipy \citep[][www.scipy.org]{scipy} and IPython \citep{Perez}. Plots have been produced with matplotlib \citep{Hunter}.\\
This research made use of Astropy, a community-developed core Python package for Astronomy \citep{astropy}\\
This research made use of APLpy, an open-source plotting package for Python \citep{aplpy}\\

\end{acknowledgements}

\bibliographystyle{B20924+30.bst}
\bibliography{B20924+30.bib}

\end{document}